\documentclass[acmsmall, screen]{acmart}

\pdfoutput=1

\AtBeginDocument{%
  }

\usepackage{microtype}
\usepackage{subfigure}
\usepackage{calc}
\usepackage{framed}
\usepackage{makecell}
\usepackage{xspace}
\usepackage{amsthm}
\usepackage{amsmath}
\usepackage{amstext}
\usepackage[dvipsnames]{xcolor}
\usepackage{graphicx}
\usepackage{enumitem}
\usepackage{url}

\newlength\nesw
\newlength\nwse
\newcommand{\ket}[1]{|{#1} \rangle}
\newcommand{\bra}[1]{{\langle {#1}|}}

\newcommand{\fzero}{0}
\newcommand{\fone}{1}

\newcommand{\spn}{\operatorname{span}}

\newcommand{\braket}[2]{\langle {#1} | {#2} \rangle}
\newcommand{\fet}[1]{| {#1} \rangle^{\!\text{\tiny F}}}

\newcommand{\calu}{U}

\newcommand{\inv}{{\operatorname{invalid}}}
\newcommand{\err}{{\operatorname{error}}}
\newcommand{\loss}{{\operatorname{loss}}}

\newcommand{\mapm}[1]{\stackrel{#1}{\longrightarrow}}

\newcommand{\sA}{\mathrm{A}}
\newcommand{\sB}{\mathrm{B}}
\newcommand{\sC}{\mathrm{C}}
\newcommand{\sE}{\mathrm{E}}
\newcommand{\sP}{\mathrm{P}}
\newcommand{\sH}{\mathrm{H}}
\newcommand{\sV}{\mathrm{V}}
\newcommand{\BS}{\mathrm{BS}}
\newcommand{\sanc}{\mathrm{anc}}

\newtheorem{theorem}{Theorem}

\newtheorem{definition}{Definition}

\newtheorem{observation}[theorem]{Observation}
\newtheorem{property}{Property}

\begin{document}

\title{Cybersecurity of Quantum Key Distribution Implementations}

\author{Ittay Alfassi}
\email{ittay.al@cs.technion.ac.il}
\orcid{0009-0007-8051-2469}
\affiliation{%
  \department{Computer Science}
  \institution{Technion---Israel Institute of Technology}
  \city{Haifa}
  \country{Israel}
}

\author{Ran Gelles}
\email{ran.gelles@biu.ac.il}
\orcid{0000-0003-3615-3239}
\affiliation{%
  \department{Faculty of Engineering}
  \institution{Bar-Ilan University}
  \city{Ramat Gan}
  \country{Israel}
  }

\author{Rotem Liss}
\email{rotem.liss@icfo.eu}
\orcid{0000-0003-1579-0095}
\affiliation{%
  \institution{ICFO---Institut de Ciencies Fotoniques, Barcelona Institute of Science and Technology}
  \city{Castelldefels}
  \state{Barcelona}
  \country{Spain}
}

\author{Tal Mor}
\email{talmo@cs.technion.ac.il}
\orcid{0000-0003-2074-0498}
\affiliation{%
  \department{Computer Science}
  \institution{Technion---Israel Institute of Technology}
  \city{Haifa}
  \country{Israel}
}
\affiliation{%
  \institution{The Helen Diller Quantum Center}
  \city{Haifa}
  \country{Israel}
}

\renewcommand{\shortauthors}{Alfassi, Gelles, Liss, and Mor}

\begin{abstract}
Practical implementations of Quantum Key Distribution (QKD) often deviate from the theoretical protocols, exposing the implementations to various attacks even when the underlying (ideal) protocol is proven secure.
We present new analysis tools and methodologies for \textit{quantum cybersecurity}, adapting the concepts of vulnerabilities, attack surfaces, and exploits from classical cybersecurity to QKD implementation attacks.
We also present three additional concepts, derived from the connection between classical and quantum cybersecurity:
``Quantum Fuzzing'', which is the first tool for black-box vulnerability research on QKD implementations;
``Reversed-Space Attacks'', which are a generic exploit method using the attack surface of imperfect receivers;
and concrete quantum-mechanical definitions of ``Quantum Side-Channel Attacks'' and ``Quantum State-Channel Attacks'', meaningfully distinguishing them from each other and from other attacks.
Using our tools, we analyze multiple existing QKD attacks and show that the ``Bright Illumination'' attack could have been found even with minimal knowledge of the device implementation.
This work begins to bridge the gap between current analysis methods for experimental attacks on QKD implementations and the decades-long research in the field of classical cybersecurity, improving the practical security of QKD products and enhancing their usefulness in real-world systems.
\end{abstract}



\maketitle

\section{Introduction}
\label{sec:intro}
Quantum Key Distribution (QKD)~\cite{BB84,B92,E91,COW-QKD,Xu2020} allows two parties to generate a secret shared key
whose unconditional security is guaranteed by the principles of quantum mechanics.
In recent decades, many security proofs for QKD have been devised (e.g.,~\cite{MayersBB84Sec,ShorBB84Sec,BBBMR00_conf,BBBMR06,RennerBB84SecCompose, BenOrBB84SecCompose}), proving security under specific theoretical assumptions.

QKD schemes have not only been described theoretically, but also implemented in practice, including in laboratory experiments and commercial products~\cite{bennett_experimental_1992,muller_plug_1997,Stucki_2009,Xu2020,Peev2009}.

Unfortunately, those real-world implementations of QKD inevitably deviate from the idealized models assumed in security proofs, leading to possibly exploitable loopholes.
As a simple example, while most QKD schemes are based on qubits, they are implemented by sending and receiving photons, which actually reside in a quantum space of higher dimension.
Such mismatches between theoretical models and practical implementations of QKD have given rise to various attacks that effectively compromise the security of the secret key~\cite{PNS,vakhitov_large_2001,makarov_faked_2005,gisin_trojan-horse_2006,qi_time-shift_2006,makarov_faked_2008,lydersen_blinding_commercial_2010,boyer_fixedapp_2014,Pang2020};
these attacks highlight the need for a deeper understanding of security in \textit{practical} QKD systems and better ways to analyze their security.

In contrast to the analysis of QKD imperfections, the security analysis and certification of practical implementations of communication systems
have been thoroughly researched in the classical field of cybersecurity, resulting in well-established methods for such analyses~\cite{shellcoders_handbook,Ross2021_SP800}.
The first step in such an analysis is to review the interfaces of the system with potential adversaries, interfaces named \textit{the attack surface}.
The next step (called ``vulnerability research'') is to search for \textit{vulnerabilities}, which are unwanted behaviors of the system, some related to the protocol the system implements and some to the environment, the latter called \textit{side channels}.
The final step is to build an \textit{exploit}: an attack that uses the detected vulnerabilities.

The purpose of this paper is to build an analogous
set of tools for \textit{quantum cybersecurity}, bringing insights from classical cybersecurity into the world of QKD systems.
We begin by defining equivalents of concepts and tools from classical cybersecurity, including vulnerabilities, attack surfaces, and exploits (Section~\ref{sec:vulns_exploits}).
We then show how to decompose QKD implementation attacks into those components and analyze each component separately, enabling QKD system designers to systematically discover and prevent possible attacks.

Then, we define the first vulnerability research method for QKD implementations, which we name \textit{Quantum Fuzzing} (see Section~\ref{sec:fuzzing}). This method is analogous to the notion of classical fuzzing~\cite{fuzzing,fuzzing_review_1,fuzzing_review_2}, which ``experiments'' with unusual inputs to an implementation in order to expose vulnerabilities.
Adapting this concept to the quantum world, we enable QKD system designers and attackers to discover vulnerabilities in a QKD system, even if their knowledge of its inner workings is limited.

We next define a general
family of attacks named \textit{Reversed-Space Attacks} 
(Section~\ref{sec:RSA}),
which is a generic exploit method for discovered vulnerabilities. 
This method, initially introduced in the preliminary work of~\cite{gelles_security_2012,gelles_reversed_2016}, begins by finding the \textit{reversed space} of the QKD implementation, which is derived by reversing states that are measured by the QKD device ``back in time''. This space is essentially the receiving party's \textit{attack surface}, because it exactly describes the quantum space affecting their measurements. 
Then, the method uses the reversed space to construct a successful attack (see Appendix~\ref{app:rsa_main_example} for a detailed example). 
We note that an additional important example of a reversed-space attack is the ``fixed-apparatus attack'', which appeared in~\cite{boyer_fixedapp_2014}.

In contrast to attacks that only rely on the transmitted and received quantum states, there are also attacks that utilize 
unmeasured side channels; such attacks are named ``side-channel attacks''.
Side-channel attacks on quantum devices have been discussed in the literature~\cite{makarov_faked_2008,lydersen_blinding_commercial_2010,after_gate,bsi} as attacks that rely on physical faults; however, these previous definitions were too general, because \textit{all} attacks on QKD systems rely on physical faults.
We thus give new and meaningful definitions of \textit{Quantum Side-Channel Attacks} and \textit{Quantum State-Channel Attacks} (Section~\ref{sec:side_channels}). 

To demonstrate how the methods we developed can help devise and analyze attacks on practical QKD implementations, we show that the \textit{Bright Illumination attack}~\cite{makarov_blinding_first_2009,lydersen_blinding_commercial_2010,liss_practice_2020} is, in fact, a delicately crafted exploitation of the tools defined in this paper (Section~\ref{sec:bi_analysis}).
Finally, we use our tools to classify well-known attacks on practical QKD implementations (Section~\ref{sec:applicability}).

\smallskip
Initial versions of (subsets of) the results presented in this paper previously appeared in~\cite{gelles_security_2012} (see also extended results in~\cite{gelles_reversed_2016}) and~\cite{liss_practice_2020}.

\subsection{Other Related Work}
Our work focuses on implementations of QKD schemes using \textit{photons}.
This is not a limitation, since all commercial implementations and almost all experimental implementations of QKD are photon-based~\cite{bennett_experimental_1992,muller_plug_1997,Stucki_2009,Xu2020,Peev2009,Yuan2018,Zhang2020,Chen2021,Kwek2021,BassoBasset2021}.

Many attacks on QKD implementations have appeared in the literature (see, e.g.,~\cite{PNS,vakhitov_large_2001,gisin_trojan-horse_2006,lydersen_blinding_commercial_2010,Pang2020,qi_time-shift_2006,boyer_fixedapp_2014}; see Appendix~\ref{app:attack_list} for more details).
Previous works have also classified certain similar attacks together as ``attack families'', which are sets of attacks that share common principles; two examples include the ``faked states'' attack family~\cite{makarov_faked_2005,makarov_faked_2008} and the ``Detector Efficiency Mismatch'' attack family~\cite{makarov_effects_2006}.

A thorough survey of attacks on QKD implementations was given in~\cite{bsi}. In addition, the term ``vulnerabilities'' in reference to QKD imperfections was also used in the review~\cite{Zapatero2025}.

Finally, in this paper, we use classical cybersecurity terminology that has been commonplace in the last few decades.
See~\cite{shellcoders_handbook} for an educational handbook and the NIST Special Publication 800-160 documents~\cite{Ross2021_SP800} for cybersecurity guidelines of the US federal government: both provide documentation of that terminology, as well as a broad introduction to classical cybersecurity as a whole.

\section{Preliminaries}
\label{sec:prelim}

\subsection{The (Ideal) BB84 Protocol}
The first and most prominent QKD protocol is BB84,
invented by Bennett and Brassard~\cite{BB84}.
It allows two legitimate participants,
typically named Alice and Bob, to generate a secret shared key.

The protocol begins with a quantum communication phase,
where Alice sends $N$ transmissions to Bob.
In each transmission, Alice randomly chooses a quantum state
out of the set $\{\ket{\fzero}, \ket{\fone},
\ket{+} \triangleq \frac{\ket{\fzero} + \ket{\fone}}{\sqrt{2}},
\ket{-} \triangleq \frac{\ket{\fzero} - \ket{\fone}}{\sqrt{2}}\}$,
and Bob randomly chooses whether to measure in the
computational basis $\sB_\sC = \{\ket{\fzero}, \ket{\fone}\}$
or the Hadamard basis $\sB_\sH = \{\ket{+}, \ket{-}\}$.
Then, Alice transmits her chosen state,
and Bob measures it in his chosen basis.

After all transmissions have been completed,
Alice and Bob continue the protocol using
a classical authenticated communication channel
(to which where Eve can listen, but not interfere).
First, Alice reveals whether each of her transmissions
was made in the computational or the Hadamard basis,
and Bob reveals which basis he used
for measuring the transmission.
Alice and Bob then discard all transmissions where they used
mismatching bases (since the results carry no information)
and retain a set of measurement results
that should be equal if no eavesdropping or noise occurred.

Thereafter, Alice and Bob choose a random subset of their
measurement results and publish it over the classical channel,
estimating the amount of mismatches in their shared string.
The fraction of errors in transmission
(that is, the number of transmissions that had errors,
divided by the total number of transmissions)
is called the quantum bit error rate (QBER).
If the QBER is above a certain threshold,
an eavesdropper has completely compromised
the security of the transmissions, and the protocol is aborted;
otherwise, Alice and Bob continue the protocol with
the non-published secret bits, defined as the raw key.
Since the raw key can still include mismatches
between Alice and Bob, and since Eve can
still have little information about it,
Alice and Bob perform classical phases of
error correction and privacy amplification
to transform the raw key into a shorter string (the final key)
which is completely equal for Alice and Bob
and (up to a negligible probability)
completely secret from Eve.

\subsection{Realistic QKD Implementations and Fock Space Notation}
\label{subsec:fock_notation}
While the ideal BB84 (and other QKD protocols) communicates using abstract qubits, 
any experimental implementation must rely on physical, easy-to-transfer quantum carriers, which encode qubits.
In most existing implementations, \textit{photons} serve as such carriers.
A very well-known example (e.g.,~\cite{bennett_experimental_1992}) is the  BB84 implementation that uses two polarization modes to encode a qubit: namely, the polarizations
$\leftrightarrow$~and~$\,\updownarrow$ signify $\ket{0}$ and $\ket{1}$ respectively, and the orthogonal diagonal photon polarizations encode  $\ket{+}$ and $\ket{-}$.

\paragraph*{Fock States.}
To accurately describe photonic quantum systems, one can use Fock Space notation (or simply, Fock states).
Photons are physical systems with many quantum degrees of freedom, called \textit{modes}, that describe different characteristics of the photons such as their polarization, location, etc.

Specifically, the Fock state $\fet{m}$
represents $m$ identical photons in a single photonic mode.\footnote{We use the notation $\fet{\cdot}$ to indicate use of the occupancy number basis.}
If the system uses photons with different characteristics,
they are represented by different modes.
In particular, the state
$\fet{m_1} \otimes \fet{m_0} \equiv \fet{m_1}\fet{m_0} \equiv \fet{m_1,m_0}$ describes $m_0$ photons in the first mode
and $m_1$ photons in the second mode. This can be extended to any number of modes and photons in each mode.

Using this notation, the above polarization-based implementation can be expressed using two modes: one mode signifying photons in the horizontal polarization (denoted $\leftrightarrow$ or $\sH$) and another mode signifying photons in the vertical polarization (denoted $\updownarrow$ or $\sV$).
The state $\ket{0}= \fet{0{,}1}_{\sV,\sH}$ describes a single photon with horizontal polarization, and $\ket{1}=\fet{1{,}0}_{\sV,\sH}$ describes a single photon with vertical polarization. 
Similarly, in implementations that encode a qubit onto the \textit{time} at which a photon arrives, we can define two distinct time bins:
$t_0$ and $t_1$. In this case, $\ket{\fzero} = \fet{0{,}1}_{t_1,t_0}$ is a photon arriving at the first time-bin, and $\ket{\fone} = \fet{1{,}0}_{t_1,t_0}$ is a single photon arriving at the second time-bin.
We will use $\ket{V}=\fet{0{,}0}$ to denote the \textit{vacuum} state --- a state with zero photons in all modes.
Note that since photons are indistinguishable and have integer spin, the state remains identical when photons ``switch places'', and there is no upper bound on the number of photons in each mode.

\paragraph*{Realistic Measurement of Photons.}
Theoretically, measurement of the Fock state 
$\fet{m}$ can yield the number of photons occupying the mode --- that is, the number~$m$.
This can be extended to an ideal measurement of the $k$-mode Fock state
$\fet{m_{k-1},\ldots, m_1,m_0}$ which yields the 
numbers $m_0$ to~$m_{k-1}$. 
However, as shown in the following sections, realistic devices are typically unable to count the exact number of photons.

In addition, one can measure other specific properties of the state 
using linear optical tools, such as beam splitters, phase 
shifters, and mirrors (see~\cite{RZBB94}). Details on the applications of these tools
on photonic quantum states can be found in Appendix~\ref{app:quantum_optics}.

\section{Classical and Quantum Cybersecurity}
\label{sec:vulns_exploits}
The security of classical computing systems, including the structure of potential weaknesses and methods for their analysis, has been extensively researched in the field of classical cybersecurity.
In this section we explain the concepts and methodology
used for analyzing computing systems via classical cybersecurity, and we explain how they can be utilized and applied to the world of QKD implementations.

\subsection{Classical Cybersecurity}
The concepts of classical cybersecurity are best described using an example: we consider a blog website running on a server.
A security researcher can analyze the security of that blog using the following method: 

\begin{enumerate}
    \item \textbf{Examining the Attack Surface.}
    The first step in analyzing a system’s security is to examine the various interfaces through which an attacker could interact with the system. 
    In a standard website-based blog, an attacker can target any interface normally available to all blog readers. For example, a reader can typically post a comment by sending a ``post comment'' request. 
    Sometimes, there are additional available interfaces.
    For instance, there might be a way to send and receive files from the blog-hosting server (such as an FTP interface), or even access the blog administrator's interface. 
    These interfaces may be blocked to external connections, protected by passwords, or left completely open. 
    Other potential interfaces, such as physical access to the server (for example, plugging in an infected USB drive), are not available to an attacker outside the premises, but could be available to an internal attacker, such as a malicious employee.

    The collection of interfaces accessible to an attacker is called the \textit{attack surface} of the system. 
    \begin{definition}
    \label{def:attackSurface}
        An attack surface is ``the set of points on the boundary of a system, a system element, or an environment where an attacker can try to enter, cause an effect on, or extract data from''.~\cite{Ross2021_SP800}
    \end{definition}
    Note that this is a broad definition, which means that when analyzing a system’s attack surface, one must take into account all possible interfaces an attacker has with the system and all potential effects those interfaces can have on it.

    \item \textbf{Detecting and Analyzing Vulnerabilities.}
    Once our security researcher examines the attack surface of the blog, they can try finding \textit{loopholes}, 
    i.e., flaws in the implementation of the blog, that may be exploited in order to gain unauthorized access to some functionality or resource of the system.
    Such imperfections are called \textit{vulnerabilities}. 
    \begin{definition}
    \label{def:vulnerability}
        A vulnerability is ``a flaw in a system's security that can lead to an attacker utilizing the system in a manner other than that which the designer intended''.~\cite{shellcoders_handbook}
    \end{definition}
    Naturally, a security researcher should focus on vulnerabilities that are accessible through the system’s attack surface. 
    Other vulnerabilities, even if they exist and could enable harmful attacks, are irrelevant to developing an exploit by the considered attacker.

    An example of a well-known vulnerability is a ``Buffer Overflow'' (BOF) (see, e.g.,~\cite{aleph1996smashing}), where a program writes data to an array of bytes (a buffer) that is shorter than the data length.
    This causes the data to overflow from the confines of the dedicated buffer to the next memory cells, affecting other data in the computer's memory.
    In the context of our blog website, such a vulnerability may occur in the code handling the ``post comment'' request if a user can send, for example, an arbitrarily long email address while the server allocates a fixed-size buffer to store it.
    
    The vulnerabilities mentioned above are based on direct interaction of the user (or the attacker) with the system.
    In contrast, there are vulnerabilities that 
    rely on indirect interaction of the attacker with the system, extending beyond standard inputs and outputs. 
    Attacks based on this different type of vulnerability are known as \textit{side-channel attacks}.
    As a toy example, consider the case where the blog verifies a user's password by comparing it one character at a time to the stored password (instead of storing a hash of the password), and the server does not limit the amount of login attempts.
    An attacker could measure the blog's response time to infer  how many characters were matched correctly.

\item \textbf{Developing Exploits.}
The mere existence of a vulnerability, even if it causes some undesired behavior, is not sufficient by itself to compromise a system’s security. 
To do so, the vulnerability must be triggered as part of a well-planned \textit{attack}, which can produce the desired effect on the target system.
For example, if an attacker finds out that the blog has a Buffer Overflow vulnerability when processing an email input, they can submit a well-crafted email-like input whose overflowing part would make the blog's server run some malicious code.

An attack that utilizes a vulnerability in order to gain unauthorized access to a system or its data, or to maliciously affect them, is called an \textit{exploit} (as a noun).
\begin{definition}
\label{def:exploit}
    An exploit is a ``tool, set of instructions, or code that is used to take advantage of a vulnerability''.~\cite{shellcoders_handbook}.
\end{definition}
\end{enumerate}

The above concepts are highly beneficial for analyzing attacks on computing systems because of their ability to \textit{decompose attacks into separate stages and components}.
Once each component is considered separately, researchers can better understand how this component behaves, find more instances of it, and build tools to defend against it.
For example, researching vulnerabilities has given rise to definitions of common types of vulnerabilities~\cite{aleph1996smashing}, methods for discovering vulnerabilities in code~\cite{fuzzing}, and coding practices that minimize vulnerabilities~\cite{seacord2013secure}.
Researching attack surfaces has given rise to attack surface characterization methods~\cite{Manadhata_metric_2011}, as well as tools for minimizing attack surfaces (such as firewalls).
Researching the concept of exploits has given rise to generic exploit methods~\cite{shacham2007geometry}, as well as defense mechanisms that prevent exploits~\cite{shacham2004effective,abadi2005cfi}.

\subsection{Cybersecurity of QKD systems}
We can now analyze the security of QKD systems through the lens of the classical cybersecurity analysis framework depicted above, extending the framework to QKD systems.
This allows us to decompose complete attacks into their constituent components and analyze each one in detail, as is standard practice in classical cybersecurity.

\subsubsection{Vulnerabilities}
\label{subsubsec:qkd_vulnerabilities}
Vulnerabilities in QKD systems are certain ``imperfections'' and ``weaknesses'' that occur in practical implementations when they deviate from the model defined by the theoretical protocol.
We will now give examples of such imperfections, both in Alice's state preparation devices and in Bob's measurement devices.

\paragraph*{Vulnerabilities in Alice's Preparation Devices: Photon-Number Splitting}

A common form of vulnerability in Alice's device is that Alice sometimes sends different states than those she intended. A well-known example of such a vulnerability can be seen in the \textit{Photon-Number Splitting} (PNS) attack~\cite{PNS, PNS_eurocrypt} (which showed all QKD experiments done until around year 2000 to be insecure).
The PNS vulnerability is based on a fault in Alice's transmitter that sometimes sends \textit{two-photon pulses} (instead of a single photon) in one of the four allowed configurations.

Eve can build an exploit based on this vulnerability by distinguishing the two-photon pulses from the single-photon pulses, keeping one photon for herself, then measuring it once the classical bases are revealed.

\paragraph*{Vulnerabilities in Bob's Detectors}
A common vulnerability in Bob’s detectors is that they measure a Hilbert space larger than an ideal qubit. This enlargement can arise either from flaws in the physical components or from deliberate expansions of the measurement space by Bob’s devices. We now present two examples of such vulnerabilities caused by physical imperfections.

\subparagraph*{Example 1: Photon Number Resolution}
\label{subsec:counting_vuln}
Suppose Alice sends a perfect qubit encoded into the polarization of a single photon, and yet Bob uses a detector that cannot distinguish a single photon from a pair of photons (commonly called a ``threshold detector''). 
Thus, while Alice only sends states from the Hilbert space $\spn\{\fet{0{,}1}, \fet{1{,}0}\}$, if the state $\fet{0{,}2}$ arrives at Bob's detector, the detector cannot distinguish it from the state $\fet{0{,}1}$.
Thus, Eve can attack a larger space than is possible in the ideal case.\footnote{There are known ways to mitigate this vulnerability, such as using the squashing model~\cite{squashing}.}

\subparagraph*{Example 2: Detection Timing} 
\label{subsec:timing_vuln}
Suppose Alice sends a perfect qubit encoded to the polarization of a single photon, which arrives at a specific time $t$, and yet Bob cannot exactly distinguish \textit{when} a photon arrives.
Namely, for some delay $\delta > 0$, he treats late photons that arrive at $t+\delta$ the same as photons that arrive correctly at time $t$.

We model this example using Fock states, with four modes describing ``horizontal at $t$'' ($\sH_t$), ``vertical at $t$'' ($\sV_t$), ``horizontal at $t+\delta$'' ($\sH_{t+\delta}$), and  ``vertical at $t+\delta$'' ($\sV_{t+\delta}$): the most general Fock state is denoted by $\fet{m_3,m_2,m_1,m_0}_{\sV_{t+\delta}, \sH_{t+\delta}, \sV_t, \sH_t}$. Alice's qubit is encoded using the span of $\{\fet{0{,}0{,}0{,}1}, \fet{0{,}0{,}1{,}0}\}$, but Bob interprets $\fet{0{,}1{,}0{,}0}$ the same as $\fet{0{,}0{,}0{,}1}$ and interprets $\fet{1{,}0{,}0{,}0}$ the same as $\fet{0{,}0{,}1{,}0}$, because Bob can properly identify the polarization ($\sH$ or $\sV$) but cannot distinguish the time $t$ from $t+\delta$.

\paragraph{Examples of Vulnerability Classes in QKD}
Taking inspiration from the examples above, we define two classes of vulnerabilities in Bob's system that will serve us throughout the paper.

The first class of vulnerabilities describes a deviation from the theoretical protocol that almost always occurs: Bob's detectors measure more than the space on which the protocol states are encoded. Thus, the space Bob measures is larger than it has to be.
\begin{definition}
\label{def:msv}
	A \emph{Measurement Space Vulnerability} occurs when Bob's measured space differs from the space defined in the theoretical protocol.
\end{definition}
An example of a Measurement Space Vulnerability is a detector that measures pulses at time $t$ and at time $t+\delta$, when it should only measure at time~$t$, as seen in Example~2 above.

\smallskip

A Measurement Space Vulnerability is often not exploitable into a full attack by itself, since there is another critical element in Bob's system: the \textit{interpretation} of the new, unexpected states.
When Bob measures a different space than he should, he might also ``confuse'' states that belong in the ideal space with states that do not.
We now define a second class of vulnerabilities, named \textit{Interpretation Vulnerabilities}, that describe this phenomenon.

\begin{definition}
\label{def:iv}
    An \emph{Interpretation Vulnerability} occurs when Bob's interpretation of measured states as valid states differs from the theoretical protocol.
\end{definition}
Example~2 above also illustrates an Interpretation Vulnerability:
Bob does not only measure pulses at $t+\delta$, but also interprets the time-shifted states as if they were not shifted.
Thus, 
Example~2 can be decomposed into two parts: 
a Measurement Space Vulnerability, in which additional states are measured, and an Interpretation Vulnerability, in which those states are misinterpreted.
This holds in general: an Interpretation Vulnerability must always be accompanied by a Measurement Space Vulnerability.

Additional interesting properties of these vulnerability classes will be discussed in the following subsections.

\subsubsection{Attack Surface}
\label{subsubsec:qkd_attack_surface}
The attack surface of a QKD implementation can be divided into two components.
The first includes interfaces defined by state spaces the devices use. The second regards other elements in the devices that interact with the environment, and by extension, with the attacker. Let us describe these two in turn.

\paragraph*{The state spaces in use.}
Every QKD device, be it Alice's or Bob's, uses a specific state space for its operation. Each state space is a component of the system's attack surface.
In Alice's device, the state space component of the attack surface is the span of the states she transmits.
For example, in the ideal BB84, Alice transmits qubits.
However, implementation vulnerabilities can enlarge this space: for example, if Alice's device suffers from the Photon Number Splitting (PNS) vulnerability, her attack surface changes from the space spanned by $\{\fet{0{,}1},\fet{1{,}0}\}$ to the space spanned by $\{\fet{0{,}0},\fet{0{,}1}, \fet{1{,}0}, \fet{0{,}2}, \fet{1{,}1}, \fet{2{,}0}\}$.

In Bob's device, the state space component of the attack surface is the space that affects Bob's measurement.
For example, in the ideal BB84, this is again the qubit space. However, 
if Bob's device suffers from the detection timing vulnerability stated in Example~2, instead of being affected by the space spanned by $\{\fet{0{,}0{,}0{,}1},\fet{0{,}0{,}1{,}0}\}$, he will be affected by the span of $\{\fet{0{,}0{,}0{,}1},\fet{0{,}0{,}1{,}0}, \fet{0{,}1{,}0{,}0}, \fet{1{,}0{,}0{,}0}\}$.
The connection to Measurement Space Vulnerabilities (Definition~\ref{def:msv}) is clear: if a device suffers from a Measurement Space Vulnerability, the space it measures is different from the ideal space, which modifies the attack surface and usually enlarges it, as occurs in the above example.

\paragraph*{Elements that interact with the environment.}
Every device interacts with its physical environment to a certain degree. Depending on how drastically a QKD device can be affected through its interaction with the environment, or how much internal information is exposed through it, that interaction can enable critical attacks on the QKD device.
For example, consider the physical configuration of Alice's or Bob's device when they prepare to send or measure a state.
If this configuration can somehow be probed by Eve, then it is part of the attack surface of the devices, and Eve can potentially use it for an attack~\cite{vakhitov_large_2001}.
Section~\ref{sec:side_channels} defines attacks based on this part of the attack surface as ``Quantum Side-Channel Attacks'' and includes a detailed explanation of the above example.

\subsubsection{Exploits}
\label{subsubsec:qkd_exploits}
Similarly to the classical-world definition, an exploit is an attack that utilizes a vulnerability in the system.
In QKD implementations,
such an attack is usually modeled as a single unitary transformation that Eve can apply, although in the most general case it can be modeled as a multi-stage procedure.

Since all successful attacks on QKD must be based on some imperfection in the implementation, every successful attack can be partitioned into two parts: a vulnerability (or several vulnerabilities) and an exploit that utilizes it.
This decomposition of all attacks to vulnerabilities and exploits highlights that, just like in classical systems, finding a vulnerability in a QKD system is \textit{not enough} to attack it. To carry out a successful attack, the vulnerability must be used through a matching exploit.
From this, we conclude that building exploits is a field of its own that requires careful consideration; in fact, even when a vulnerability (or several) is already known, building an exploit is a completely separate task that can be very challenging.

\paragraph{Generic exploit methods.}
A common theme in exploit research for classical cybersecurity is the idea of generic exploit methods that apply to \textit{classes} of vulnerabilities, and not necessarily to one specific vulnerability.
Assuming that a class of vulnerabilities triggers similar behavior 
in a system (often called a \textit{primitive} in classical cybersecurity), one can build a generic exploit method that relies on that behavior.
This idea is also commonplace in existing QKD literature discussing ``attack families''. 

One known example is the ``faked states'' attack family~\cite{makarov_faked_2005}.
Attacks in this family share one common property: Eve receives the state that Alice sent, measures it, and then re-sends a state that will force a \textit{specific}
interpretation in Bob's system. 
Using our new methodology, this attack family can be decomposed to a set of vulnerabilities where fake, out-of-protocol states can force a specific interpretation, and a single, generic exploit method (which is discussed below).

Another known example is the \textit{detector efficiency mismatch} attack family~\cite{makarov_faked_2008}, which is a special case of the \textit{faked states} attack family.
Attacks in this family rely on vulnerabilities where different detectors are more/less sensitive depending on some characteristic of the input states. The generic exploit of this attack family creates attacks that manipulate this characteristic, such that whenever Eve and Bob choose different bases, Bob is less likely to detect the modified signal.

Section~\ref{sec:RSA} shows a novel attack family called \textit{Reversed-Space Attacks}, which is a generic exploit method for the Measurement Space Vulnerability and Interpretation Vulnerability classes defined in Section~\ref{subsubsec:qkd_vulnerabilities} above.

\paragraph{Example: Faked States exploit.}
We now explicitly show the generic exploit that all faked states attacks are based on, in the form of unitary transformations.
As stated above, all faked-states attacks are based on the same type of vulnerability:
there is a state that Eve can send, which forces a specific bit and basis interpretation in Bob's system.

In this exploit, Eve begins by choosing either the computational or the Hadamard basis (for each transmission). Then, Eve's unitary transformation changes each basis state from her chosen basis into its ``faked'' counterpart, which will then force the desired bit and basis detection in Bob's system. Eve's unitary also stores the value of her sent faked state in her own ancilla register, which she then measures.

Let us denote a BB84 state entering Bob's system as $\ket{\psi}_{\sB}$, and its faked counterpart as $\ket{\psi_\mathrm{fake}}_{\sB}$.
Let the vectors $\ket{E_0}_{\sE}^{\sanc}, \ket{E_1}_{\sE}^{\sanc},\ket{E_2}_{\sE}^{\sanc}, \ket{E_3}_{\sE}^{\sanc}$ be four orthogonal state vectors, and let $U_{\sE}^{\mathrm{c}}$ and $U_{\sE}^{\mathrm{H}}$ be Eve's unitary operations (depending on her basis choice).

A faked state attack on a BB84 implementation is of the form:
\begin{align}
    \begin{array}{ccc}
         U_{\sE}^{c} \ket{0}_{\sA} \ket{0}_{\sE}^{\sanc} &=& \ket{0_\mathrm{fake}}_{\sB} \ket{E_0}_{\sE}^{\sanc} ~ , \\
    U_{\sE}^{c} \ket{1}_{\sA} \ket{0}_{\sE}^{\sanc} &=& \ket{1_\mathrm{fake}}_{\sB} \ket{E_1}_{\sE}^{\sanc} ~ ,
    \end{array}
    & &
    \begin{array}{ccc}
    U_{\sE}^{H} \ket{+}_{\sA} \ket{0}_{\sE}^{\sanc} &=& \ket{+_\mathrm{fake}}_{\sB} \ket{E_2}_{\sE}^{\sanc} ~ , \\
    U_{\sE}^{H} \ket{-}_{\sA} \ket{0}_{\sE}^{\sanc} &=& \ket{-_\mathrm{fake}}_{\sB} \ket{E_3}_{\sE}^{\sanc} ~ ,
    \end{array}
\end{align}

This example illustrates the distinction between the specific vulnerability of Bob's device, and the generic exploit method which can be applied to any vulnerability from the type we analyze. It thus shows the importance and usefulness of separating the process of finding vulnerabilities (commonly referred to as ``vulnerability research'') from the process of devising an exploit, whether it is a tailor-made exploit for one specific vulnerability, or a generic exploit method for a set of vulnerabilities.

\section{Quantum Fuzzing: Vulnerability Research for QKD}
\label{sec:fuzzing}
In this section, we define ``Quantum Fuzzing'', which is the first systematic vulnerability research method for QKD devices.
In classical cybersecurity, vulnerability research is the process of analyzing a device's implementation in order to detect currently unknown vulnerabilities and issues.
While it is quite common for classical system designers and attackers to perform, it has yet to be applied to QKD, where, in theory, all devices are provably secure, and all adversaries are aware of all possible attacks on a device.
We show that this process becomes crucial when considering practical QKD implementations and adversaries.
Our method, which relies on minimal knowledge of a device's inner workings, enables system designers to detect issues that would otherwise require expert knowledge to be identified, or worse, would only be discovered when it is too late.

\subsection{Motivation: Practical Adversaries for QKD Systems}

QKD protocols aim to provide unconditional security: the generated key should be secure without making any assumption on the capabilities or knowledge of the attacker.
Similarly, when examining the security of a practical implementation, a desired goal is to guarantee the security of the implementation without imposing any conditions or limitations on Eve. 
In particular, Eve is assumed to be all-powerful and all-knowing, where her knowledge includes the structure of the implementation, and any action Alice and Bob can take as part of the protocol.
Under this assumption, we must assume that Eve has knowledge of any vulnerability that exists in the devices used by Alice and Bob, which she can and will exploit in order to conduct her attack.

In contrast, classical cybersecurity considers a wide range of possible adversaries when examining system security.
Some (resourceful) adversaries have intimate knowledge of the system's inner workings, while others do not.
Furthermore, even adversaries who know the inner workings of a system may not have detected all its vulnerabilities; the same may be true even for the system designers themselves.
Thus, we conclude that many \textit{practical adversaries} and system designers do not know all the vulnerabilities in a specified system a priori.

When an adversary wants to attack a system but does not know of any vulnerability that could be exploited, they can perform \textit{vulnerability research}---a systematic process for identifying existing vulnerabilities.
Vulnerability research can be carried out via multiple methods. 
Some require full knowledge of the internal implementation (such as source code), while others can be applied in a \textit{black-box} manner,
independent of the specifics of the implementation.

The same applies when analyzing the security of QKD systems \textit{in practice}: system designers and attackers might not be fully aware of vulnerabilities in a given implementation, making vulnerability research an important tool for identifying vulnerabilities in QKD systems.

In this section, we propose \textit{Quantum Fuzzing}, a simple and effective method for researching vulnerabilities in practical QKD systems.

\subsection{Quantum Fuzzing: Definition}
The concept of \textit{fuzzing}~\cite{fuzzing, shellcoders_handbook} is a simple yet effective method to conduct vulnerability research on a device, whether classical or quantum, by probing it with various inputs (both valid and invalid) and analyzing the results: how each input affects the device, as well as the device's output.
Adapting this concept to QKD systems, we define Quantum Fuzzing in the following way:

\begin{definition}
    \emph{Quantum Fuzzing} is the process of testing a quantum device by sending many quantum and classical states,
    studying both their effect on the device and the device's output.
\end{definition}

The concept of Quantum Fuzzing has previously been discussed in the context of quantum software testing~\cite{quanfuzz, quantfuzz2}.
However, here we wish to consider it in a different context: the one of cybersecurity and vulnerability research for QKD.

\subsection{Choosing Input States}
\label{subsec:fuzz_strategy}
All fuzzing algorithms and devices (commonly called ``fuzzers'') need an underlying strategy: what test cases should be tried, and which cases should be tried first? Should these choices change based on the device's output? If so, how?

In classical fuzzing, the subject of fuzzing strategies has been thoroughly researched~\cite{fuzzing_review_1,fuzzing_review_2}.
Various rules and statistics are used to determine which inputs are likely to yield valuable outcomes and should thus be tested first.
In addition, some classical fuzzers use the tested device's output to improve their test cases mid-run: if a certain input triggers an interesting behavior, it may be beneficial to test other similar inputs, or otherwise to concatenate this interesting input to the next test cases.

Clearly, a fuzzing strategy is also required for Quantum Fuzzing:
there are infinitely many states that can be sent to a device, and one must decide which input states to test and in what order. This strategy may be fixed or adaptive, changing the next input based on the device's responses so far.

An intuitive strategy for fuzzing QKD devices is to begin with valid input states and gradually modify them: starting from simple variations in single degrees of freedom, and progressively moving to more complex modifications.
The first test cases should be the valid protocol states, in order to verify that the device responds to them appropriately.
Then, the next test cases should add small variations to the valid states.
Good candidates for these variations can be inspired by previous, well-known attacks: shifting the frequency of the photons~\cite{tans_attack}, changing their arrival time~\cite{qi_time-shift_2006,gelles_security_2012}, changing the signal intensity~\cite{gottesman_security_2004,lydersen_blinding_commercial_2010}, etc.
If small variations do not trigger an interesting behavior in the device, the next test cases should perform more complex variations and modify more than one degree of freedom in each test case: for example, changing both the time and frequency of a photon, or testing a broader range of pulse intensities.
As test cases progress, one can increase the complexity of modifications to each degree of freedom and combine modifications to more degrees of freedom.
If a test case triggers an ``interesting'' behavior in the device, it should be documented, and potentially combined with other test cases in order to generate more interesting behaviors.

We leave the task of extending this strategy, as well as the task of constructing a physical device that can implement it, for future research.

\subsection{Advantages and Limitations of Quantum Fuzzing}
The most significant advantage of Quantum Fuzzing is its ability to reveal vulnerabilities without requiring expert knowledge: it can be applied with minimal knowledge of the target device (it only requires the attacker to know the valid input states it is supposed to handle) and can reveal the existence of vulnerabilities that would otherwise require extensive and intimate knowledge of the device to detect.
For example, in Section~\ref{sec:bi_analysis} we show how the vulnerability that underlies the Bright Illumination attack~\cite{lydersen_blinding_commercial_2010, sague_blinding_pulsed_2011} could have been found using Quantum Fuzzing.

However, Quantum Fuzzing is not without limitations.
The first limitation is its partial coverage:
any method that tests a finite number of states cannot guarantee secure behavior under an infinite number of possible input states. Thus, Quantum Fuzzing can help find issues with an implementation, but cannot prove security or guarantee the absence of vulnerabilities.

The second limitation is the imprecise knowledge of the detected vulnerability.
Some vulnerabilities occur in internal parts of an implementation. 
While Quantum Fuzzing can send an input state that will trigger these vulnerabilities and create some phenomenon that the attacker can witness,
the attacker may be unable to deduce which internal part of the device is faulty and what the precise vulnerability is, especially if they lack knowledge of the inner workings of the device.

These two limitations are also true for classical fuzzing.
First, since it is impossible to test a generic program's behavior under all possible inputs, classical fuzzing also cannot guarantee security.
Second, classical fuzzing does not always reveal full details on the detected vulnerability: for example, the mere knowledge that sending a certain message caused a website to crash does not reveal which component of the website caused the crash, the specific vulnerability in the component, or how it could be further exploited.

However, both in classical and quantum systems, even imprecise knowledge of a vulnerability holds great value: it highlights specific aspects and sub-components of the device that should be analyzed more carefully to fully characterize the detected vulnerability.

\section{Reversed-Space Attacks}
\label{sec:RSA}

At its essence, the Reversed-Space method characterizes the quantum space that affects the measurements in a QKD scheme. 
Towards this end, the method analyzes both the measurement done during the QKD scheme as well as their \textit{interpretations} by the parties. By reversing each measured state \textit{backwards} through the implementation, we can obtain the set of states that, if sent to the device, will be measured according to some given interpretation. 
This forms the attack surface (Definition~\ref{def:attackSurface}) of the QKD implementation---the possible ways to attack it. By analyzing this attack surface, i.e., the resulting reversed space, we can define possible attacks (exploits, Definition~\ref{def:exploit}) on the implementation and analyze their effectivity.

\subsection{Reversed Space Formalism: a Guiding Example}
\label{subsec:RSA_guidingExample}
Consider a photonic BB84 implementation where Alice sends (via a single pulse) a perfect qubit into the polarization of a single photon (as discussed in Section~\ref{subsec:fock_notation}).
Further suppose that Bob uses a detector that cannot distinguish a single photon from a pair of photons. 
Then, even though a pair of photons never arrives from Alice in that single pulse (since Alice is assumed to be ideal),
the quantum state describing that option could arrive at Bob from an imperfect channel
or a channel controlled by Eve, hence, ought to be taken into account. 

We now explain the scenario through the lens of the Reversed-Space formalism.
Let $\{\ket{j}_{\sB}\}_j=\{\fet{0{,}1},\fet{1{,}0}\}$ be the computational basis for  Bob's system. 
Assume without loss of generality
that Bob's device is described by a unitary transformation~$\calu_{\sB}$, followed by a measurement 
in Bob's computational basis.\footnote{
Any generalized measurement (POVM) can be described as adding an ancilla, measuring,
and possibly forgetting (treating several outcomes as the same outcome),
and thus is included within our formalization.}  
The unitary transformation~$\calu_{\sB}$, and thus the actual measurement that Bob performs, depends on 
a random input of Bob, which determines whether Bob measures the computational or Hadamard basis. 
Namely,
to measure in the computational basis,
Bob sets $\calu_{\sB} = {I}$ and measures the result in his computational basis.
To measure in the Hadamard basis, Bob lets the pulse go through optical devices such as a polarization rotator, which, on a single qubit, has the effect of performing
 $\calu_{\sB} = H \triangleq
\tfrac{1}{\sqrt{2}}\left (\begin{smallmatrix} 1 & \phantom{-{}}1 \\ 1 & -1
\end{smallmatrix}\right )$;
Bob then measures the resulting pulse in the computational basis. Note that $\calu_{\sB}$ is well-defined by the optical device for any arriving state, not only a single-photon ``qubit'' pulse.

As explained above, Bob actually receives a quantum space of a larger dimension, and his measurement outcomes may be richer than a single bit, as is the case when measuring a qubit. 
In particular, Bob can interpret 
his measurement outcome in multiple ways: (i) \textbf{information:} when the outcome indicates a specific value sent by Alice. This corresponds, for instance, to the case where Bob receives the state $\fet{0{,}1}$ or $\fet{0{,}2}$.
(ii) \textbf{a loss:}  when nothing is measured or when Alice did not send any value. This corresponds to receiving $\fet{0{,}0}$. 
(iii) \textbf{invalid outcome:} any other situation where the measurement outcome is considered invalid. This corresponds to the case where Bob gets the state $\fet{1{,}1}$ and both its detectors click.

Once we identify all the relevant states~$\ket{j}_{\sB}$ measured by Bob, and all possible transformations taken by its device,
we can apply the reversed transformation(s) $\calu_{\sB}^{-1}=\calu_{\sB}^\dag$ 
on each such state~$\ket{j}_{\sB}$. 
These states, $\{\calu_{\sB}^\dag \ket{j}_{\sB}\}$, span the space that influences Bob's outcome.
While Alice and Bob may not be aware of that enlarged space, any security analysis must assume that the attacker is fully aware of it (even if this space is not fully available to an attacker).
Indeed,
the fact that Bob's equipment measures and returns a valid result for the state 
$\fet{0{,}2}$ may be unknown to Alice and Bob, but known to Eve (e.g., due to analyzing the device or due to fuzzing).
When Eve designs her attack, she can consider all the parts of the space spanned by 
$\{\calu_{\sB}^\dag \ket{j}_{\sB}\}$ that is available to her, and she is not limited to using only ideal qubits as Alice and Bob believe they do.

\medskip

We call an attack designed according to this observation 
a \textit{Reversed-Space Attack} for a specific
reason:
the term ``reversed'' here is borrowed from the time reversal symmetry of quantum theory.
The symmetry of quantum mechanics to the exchange of the
prepared (pre-selected) state and the measured (post-selected)
state was suggested by~\cite{ABL64,AV90}, 
and was already used in quantum cryptography as well, 
see the time-reversed EPR scheme~\cite{BHM96} for example. 
Interestingly, the time-reversed EPR scheme of~\cite{BHM96} also
leads to more secure protocols, named
``measurement-device-independent QKD''~\cite{Xu2020}.

\subsection{The Reversed-Space Attack}
\label{subsec:RSA_reversed_space}
Throughout the analysis in this section we assume that Alice is ideal.
In particular, 
Alice generates and sends perfect qubits in a two-dimensional 
space $\mathcal{H}^{\sA}=\mathcal{H}_2$, where $\mathcal{H}_2 = \mathbb{C}^2$ is the 2-dimensional 
Hilbert space corresponding to a single qubit.
We denote the basis states of her system by $\ket{i}_{\sA}$ with $i \in \{0 , 1\}$.
Since Alice is ideal, her space is necessarily a subspace of the larger space that affects Bob's measuring device.

In this subsection, we formally write out the components of the Reversed-Space Attack.

\subsubsection{Bob's Measurement}
We formalize Bob's actions as
\textit{(i)} obtaining a quantum system from the channel;
\textit{(ii)} potentially adding an ancillary quantum system 
(without loss of generality, in a fixed state $\ket{0}_{\sanc}\in \mathcal{H}^{\sanc}$,
where $\mathcal{H}^{\sanc}$ is the Hilbert space of the ancillary system).
\textit{(iii)} performing a unitary transformation on the joint system
from a fixed set of $m$ possible transformations 
$\{\calu_{\sB_1}, \ldots, \calu_{\sB_m}\}$;
\textit{(iv)} measuring Bob's Hilbert space~$\mathcal{H}^{\sB}$ in a certain basis\footnote{When considering qubits,
we can assume Bob uses the computational basis, but this may not be the case in some implementations.} $B_{\sB}$. 

In order to define the reversed space relevant to this implementation, we start with Bob's possible outcomes, and we use time reversal to find the exact Hilbert space $\mathcal{H}^{\sP}$ 
which is controlled by Eve and affects Bob's measurement outcome.
First, let $\mathcal{H}^{\sB}$ be the span of  $\{\ket{j}_{\sB}\}$, for all basis states $\ket{j}_{\sB}$ measured by Bob. 
Then, $\mathcal{H}^{\sP}$ is defined to be the span of $\{\calu_{\sB_s}^\dagger \ket{j}_{\sB}\}$, 
for all $s\in \{1,\ldots,m\}$ and all basis states $\ket{j}_{\sB} \in \mathcal{H}^{\sB}$,
after tracing out any ancillary space not available to Eve 
(resulted from an ancillary space~$\mathcal{H}^{\sanc}$ added by Bob
in the ``forward in time'' description). 
Any state in Eve's space that is orthogonal to $\mathcal{H}^{\sP}$ goes, after $\calu_{\sB_s}$, to a state which is orthogonal to~$\mathcal{H}^{\sB}$ 
and can never affect Bob.

Since in the ideal-Alice case, Alice's Hilbert space~$\mathcal{H}^{\sA}$ is a subspace of~$\mathcal{H}^{\sP}$,
we treat Alice's qubit as the span of two orthonormal states in $\mathcal{H}^{\sP}$.

In the most general case, the resulting reversed space is rather complex,
as it can take into account arriving multi-photon states, arriving multi-mode states,  and potentially, the ancilla Bob's device inherently adds. 
For simplicity, one may analyze each aspect separately, although for a full security proof, one must take the combined effect into account as well. We next show how Eve's attack translates to our Reversed-Space formalism.

\subsubsection{Eve's Attack}
\label{subsec:RSA_eve_and_bob}
Since in practice Bob is affected exactly by~$\mathcal{H}^{\sP}$, Eve only needs to attack this extended space.
Thus, her most general attack can be described as adding the ancilla~$\ket{0}_{\sE}$ 
and performing a transformation~$\calu_{\sE}$ on the state $\ket{\psi}_{\sA}$ sent by Alice. Note that Alice state is actually embedded in $\mathcal{H}^{\sP}$, hence, 
Eve acts on 
$\ket{\psi}_{\sA} \rightarrow 
\ket{\psi}_{\sP} = \sum_i \alpha_i\ket{i}_{\sP} \in \mathcal{H}^{\sP}$
(where the arrow stands for an embedding). 
Eve's actions can thus be written as,
\begin{equation}\label{eqn:EvesAttack}
\ket{0}_{\sE}\ket{\psi}_{\sP} = \sum_i \alpha_i\ket{0}_{\sE}\ket{i}_{\sP} \mapm{\calu_{\sE}} 
\sum_{i,k} \alpha_i \epsilon_{i,k}\ket{E_{i,k}}_{\sE}\ket{k}_{\sP}.
\end{equation}
Thus, although Alice's (BB84) state is completely represented by the four simple options
$\{\alpha_0=1;\alpha_1=0\},
\{\alpha_0=0;\alpha_1=1\},
\{\alpha_0=1/\sqrt2;\alpha_1=1/\sqrt2\},
\{\alpha_0=1/\sqrt2;\alpha_1=-1/\sqrt2\}$, 
the state right after Eve's attack is much more complex.

Eve then sends a state in~$\mathcal{H}^{\sP}$ to Bob, who processes it as explained above. 
We can formulate Bob's action on any basis state $\ket{k}_{\sP}$,
for a given setting $\calu_{\sB_s}$with $s\in \{1,\ldots,m\}$, by
\begin{equation}\label{eqn:bob}
\ket{k}_{\sP}\ket{0}_{\sanc} \mapm{\calu_{\sB_{s}}} \sum_{j}\beta^{s}_{k,j} \ket{j}_{P\otimes\sanc},
\end{equation}
leading to the 
final state $\ket{\Psi_\mathrm{EB}}$ that Bob and Eve hold at the end of the process (just before Bob measures)
\begin{equation}
\ket{\Psi_\mathrm{EB}} \triangleq \sum_{i,k,j} \alpha_i \epsilon_{i,k} \beta^{s}_{k,j} \ket{E_{i,k}}_{\sE}\ket{j}_{P\otimes\sanc},
\label{eqn:FinalAttackedState}
\end{equation}
derived through Eqs.~\eqref{eqn:EvesAttack}--\eqref{eqn:bob},
\begin{align*}
\nonumber
&\ket{\psi}_{\sP}  \mapm{} \ket{0}_{\sE}\ket{\psi}_{\sP} \mapm{\calu_{\sE}} 
\sum_{i,k} \alpha_i\epsilon_{i,k}\ket{E_{i,k}}_{\sE}\ket{k}_{\sP} \\
&\qquad \mapm{}
\sum_{i,k} \alpha_i\epsilon_{i,k}\ket{E_{i,k}}_{\sE}\ket{k}_{\sP}\ket{0}_{\sanc} 
\mapm{\calu_{\sB_s}}
\sum_{i,k} \alpha_i\epsilon_{i,k}\ket{E_{i,k}}_{\sE}\sum_{j}\beta^{s}_{k,j} \ket{j}_{P\otimes\sanc}.\nonumber \\
\end{align*}
Finally, Bob measures the space $\mathcal{H}^{\sB}$ using his basis. Note that $\mathcal{H}^{\sB} \subseteq \mathcal{H}^{\sP}\otimes \mathcal{H}^{\sanc}$.

\subsubsection{Bob's Interpretation and Oblivious Attacks}
\label{subsec:RSA_oblivious}
In QKD implementations, the way Bob interprets his measurement outcome is of great importance. 
The states $\ket{j}_{\sB}$ can be classified into sets according to Bob's interpretation: 
some of these states indicate ``Alice has sent the bit $0$'', while others indicate ``Alice has sent the bit $1$''. 
Let us denote the set of (basis) states that Bob interprets as measuring the bit value~$0$ by~$J_0$ 
and the set of states interpreted as a~$1$ by~$J_1$.

When Alice sends a bit~$b$, but Bob measures a state in $J_{1-b}$, the transmission is said to be an \textit{error}.
Generally, for a specific transmission, we define $J_\err$ as the set of all states that Bob regards as an error.
Note that these sets are defined \textit{per transmission} and depend on the specific basis Bob uses and the state Alice
sends (i.e., the bit value $b$ she communicates).

When considering real implementations, there may be some outcomes that
are not interpreted as valid outcomes since they never happen in the ``ideal'' scheme. 
These outcomes can be divided into two groups, according to Bob's interpretation:
\begin{enumerate}
\item \textit{Outcomes interpreted as a loss:} 
transmissions that can naturally occur but supply no information to Bob. For example, vacuum pulses that make no detector click, or measurement results that are inconclusive and are part of the theoretical protocol (as occurs in the B92 protocol~\cite{B92}).
These outcomes are denoted as the set $J_\loss$.
\item \textit{Invalid-erroneous outcomes:} outcomes that can never occur if the
quantum system sent by Alice reaches Bob intact (e.g.\@ when several detectors click,
while Alice is guaranteed to send a single photon).
These outcomes are denoted as the set $J_\inv$.
\end{enumerate}
The collection of all $J_0, J_1, J_\loss$ and $J_\inv$ sets for a certain receiver implementation
are called \textit{the interpretation sets} of the receiver.

It is Bob's choice of interpretation that determines whether a
specific outcome is considered a loss or an invalid result. 
Generally speaking, when an invalid outcome increases Bob's measured 
error rate, we put it in the set~$J_\inv$, and when it is ignored by Bob, we put it in~$J_\loss$. 
As an example, let us consider the case where Alice's state generation is not totally ideal,
and she can either send a single photon or, at most, two photons. 
If Bob treats these cases of noticing many photons as a loss 
(i.e., he ignores that transmission, thus this measurement is in~$J_\loss$) rather than as an error, this results in a
major security hole~\cite{gottesman_security_2004,hwang_no-clicking_2008}.

\medskip 
We call attacks that cause no errors and no invalid outcomes at Bob's end ``oblivious''.
That is, we require that for any 
$\ket{j}$ in~$J_\err$ or in~$J_\inv$, 
the overlap~$\braket{j}{\Psi_\mathrm{EB}}$ is zero, so Bob never measures~$\ket{j}$. 
We formalize this idea using the description of the final state shown in Eq.~\eqref{eqn:FinalAttackedState}
in the following manner:
\begin{observation}\label{clm:ZeroErrorRequirement}
For a given QKD implementation, 
Eve's attack $\calu_{\sE}$ causes no errors if and only if
for every state $\ket{\psi}=\sum_i\alpha_i\ket{i}_{\sP}$ sent by Alice and for any $\calu_{\sB_s}$ used by Bob, it holds that
\begin{equation}
\label{eqn:ZeroError}
\sum_{i,k}\alpha_{i} \epsilon_{i,k}\beta^s_{k,j}
\ket{E_{i,k}}_{\sE}=0 ,
\end{equation}
for any
$j \in J_\err \cup J_\inv$  (determined according to the specific $\ket{\psi}$ sent by Alice,
and the specific setting~$s$ used by Bob).
\end{observation}

To clarify the notations for oblivious attacks, let us provide a simple example and 
show that a CNOT attack made by Eve does not satisfy the conditions of Observation~\ref{clm:ZeroErrorRequirement} and thus
can be noticed by Bob.
For instance, consider a standard BB84 scheme~\cite{BB84} in which Bob's setup 
for the computational basis is $\calu_{\sB_\sC} = I$ the identity, and for the Hadamard basis, $\calu_{\sB_\sH} = H$ is Hadamard transformation; both are followed by a measurement in the computational basis. 
Assume Alice sends~$\ket{\psi}_{\sA}=\ket{+}$, but Eve performs a CNOT attack using the computational basis.
After the attack, the system (Alice's qubit and Eve's added ancilla) is in the state
$\ket{\tilde \psi}= (\ket{E_{0,0}}_{\sE}\ket{0}_{\sP} + \ket{E_{1,1}}_{\sE}\ket{1}_{\sP})/\sqrt{2}$
with orthogonal $\ket{E_{0,0}}_{\sE}$ and $\ket{E_{1,1}}_{\sE}$.
Assume Bob sets his apparatus to the Hadamard basis (same as Alice). Thus, he applies the Hadamard transformation, and
measures subsystem $\sP$ of the resulting state
$\ket{\Psi_\mathrm{EB}}=(I_{\sE} \otimes \calu_{\sB_\sH}) \ket{\tilde \psi} = (\ket{E_{0,0}}_{\sE}\ket{+}_{\sP} + \ket{E_{1,1}}_{\sE}\ket{-}_{\sP})/\sqrt{2}$
in the computation basis.
We now show that Bob has a positive probability of measuring $\ket{1}$ (that is, $j=1$),
while this outcome is in~$J_\err$ and indicates an error:
using the formulation of Observation~\ref{clm:ZeroErrorRequirement}, 
Alice's qubit is given by $\alpha_0=\alpha_1=1/\sqrt{2}$,
Eve's attack by $\epsilon_{0,0}=\epsilon_{1,1}=1$ and $\epsilon_{0,1}=\epsilon_{1,0}=0$ 
and Bob's setup by $\beta^\sH_{0,0}=\beta^\sH_{0,1}=\beta^\sH_{1,0}=1/\sqrt{2}$ and $\beta^\sH_{1,1}=-1/\sqrt{2}$.
Indeed, Eq.~\eqref{eqn:ZeroError} for $j=1$ gives
\(
\sum_{i,k\in\{0,1\}} \alpha_i\epsilon_{i,k}\beta^\sH_{k,1} \ket{E_{i,k}}_{\sE} =
\frac1{\sqrt{2}}  %
\cdot 1 %
\cdot \frac1{\sqrt{2}}  %
\ket{E_{0,0}}
+
\frac1{\sqrt{2}}  %
\cdot 1 %
\cdot \frac{-1}{\sqrt{2}}  %
\ket{E_{1,1}}
\),
which is non-zero since $\ket{E_{0,0}}$ and $\ket{E_{1,1}}$ are orthogonal.

\medskip
Finally, we can define the set of \textit{oblivious} attacks that are ``unnoticeable'' by the parties.
\begin{definition}\label{def:Uzero}
Let $\mathbf{U}_{\rm zero}$ be the set of attacks on a given protocol,
that cause no errors (in all the possible setups of the protocol).
\end{definition}
Any attack in $\mathbf{U}_{\mathrm{zero}}$ that 
leaks some information to Eve
is considered a successful attack that potentially damages the security of the implemented QKD scheme.

\subsection{Example: a Complete Reversed-Space Attack}
\label{subsec:RSA-example}
Appendix~\ref{app:rsa_main_example} provides a detailed description of a time-based interferometric BB84 implementation, which appears in~\cite{Walton2003,Nambu2004,Jaeger2006,Sasaki2011}, and constructs a successful Reversed-Space Attack on that implementation.
This subsection gives a short overview of the implementation and the attack.

In the attacked implementation, Alice encodes her qubit using time-bin encoding with two time-bins (denoted $t'_0$ and $t'_1$).
Bob's interferometer-based device splits Alice's pulse into six separate modes: two output arms of the interferometer (denoted $s$ and $d$), where in each arm, a photon could appear in three possible time-bins (denoted $0$, $1$, and $2$). 
Thus, Bob's detectors measure \textit{six} different modes which can be affected by Alice's qubit.

A reversed-space analysis reveals that Bob's measurement is not only affected by Alice's qubit, but also by two additional times at the interferometer's input, denoted $t'_{-1}$ and $t'_2$.
The analysis of Bob's operation on these two additional times reveals that carefully-crafted pulses in the two additional times can force Bob to either measure in a basis of Eve's choice, or experience a loss of the signal.
Eve can use these additional modes to design an attack that never causes an erroneous measurement in Bob's system.

This attack breaks the security (and the robustness: see~\cite{BKM2007}) of this BB84 implementation, because Eve gains \textit{full} information on the key without inducing any error and without causing the protocol to abort.

We note that if Bob is constrained to only measure two specific modes instead of all the six resulting modes, a simpler and stronger attack can be devised; see~\cite[Section~III~D]{gelles_reversed_2016} for more details.
We also note that another successful reversed-space attack on a polarization-based interferometric BB84 implementation was reported in~\cite{boyer_fixedapp_2014}.

\subsection{Discussion: Countermeasures to Reversed-Space Attacks}
\label{subsec:rsa_defense_mechanisms}
Can Bob measure more states to defend against attacks on the enlarged space coming from Eve?
Although this seems like a natural countermeasure, it may in fact add vulnerabilities that were not originally present and open a path for new attacks: now Bob's measured space is larger, and so is its reversed space!
Hence, when analyzing the security of an implementation, one must \textit{fix} Bob's implementation before letting Eve attack it.
If we modify Bob to counter an attack, the analysis must be repeated on the new implementation, considering the new dimensions measured by Bob and his new interpretation of outcomes.

In Appendix~\ref{app:rsa_app_added_measurements}, we revisit the Reversed-Space Attack on interferometric BB84 that was described in Appendix~\ref{app:rsa_main_example}, while adding the countermeasure of measuring additional time bins in Bob's implementation.
On the one hand, our analysis shows that, under the restriction of Eve to sending single-photon pulses and non-collective attacks, Bob's given device is effective in blocking Reversed-Space Attacks.
On the other hand, we demonstrate that the additional modes measured by Bob effectively create a new implementation with a different, larger reversed space. Eve can now use a broader range of attacks than before on this new implementation.

\subsection{Summary: Reversed-Space is an Exploit}
We conclude by making the following observations, which connect the reversed-space attack to our cybersecurity analysis framework depicted in Section~\ref{sec:vulns_exploits}.

\begin{observation}
\label{claim:rs_from_msv}
    Knowing the Measurement Space Vulnerabilities of Bob's device allows the computation of the reversed space, which is a part of Bob's attack surface.
\end{observation}
By definition, Measurement Space Vulnerabilities (Definition~\ref{def:msv}) reveal the large, realistic Hilbert space that Bob measures.
The reversal of said space through Bob's setup is exactly the Hilbert space of states that Eve can send to
affect Bob.
As discussed in Section~\ref{subsubsec:qkd_attack_surface}, this is the state space part of Bob's attack surface.

\begin{observation}
    \label{claim:rs_from_iv}
    The computation of the ``zero noise requirement'' in a Reversed-Space Attack is based on Interpretation Vulnerabilities.
\end{observation}
Interpretation Vulnerabilities, by Definition~\ref{def:iv}, reveal additional information on how Bob interprets his incoming states; specifically, they reveal interpretations that should not exist in an ideal protocol, and are necessary information for Eq.~\eqref{eqn:ZeroError} that defines the ``zero noise requirement'' of Observation~\ref{clm:ZeroErrorRequirement}.
These interpretations are often the key to the success of the attack.

The following observation follows directly:
\begin{observation}
	\label{claim:rs_from_vulns}
	Reversed-Space is a generic exploit for Measurement Space Vulnerabilities and Interpretation Vulnerabilities.
\end{observation}

Reversed-Space Attacks, as an exploit method, are usable both for an attacker seeking to exploit a vulnerability (either publicly known or known only to him),
and for system designers checking the practical security of their QKD implementation.

\section{Quantum Side-Channel Attacks}
\label{sec:side_channels}

In this section, we examine and define the notion of ``side-channel attacks'' in the context of QKD devices. Similarly to the classical side-channel attacks, these attacks exploit unintended physical interfaces that potentially compromise the security of the underlying QKD protocol.
However, unlike classical side-channel attacks, physical interfaces of QKD protocols are often part of the acceptable communication between Alice and Bob.

Therefore, we introduce new definitions of ``Quantum Side-Channel Attacks'' and their complement, which we name ``Quantum State-Channel Attacks''.
We also present concrete examples to show that Quantum Side-Channel Attacks can target a wide range of device components.
Because these attacks are often overlooked in standard security analyses, our definition emphasizes the need to treat them explicitly and provides system designers with a clearer understanding and a stronger basis for defending against them.

\subsection{The Classical Concept of Side-Channel Attacks}
In cybersecurity, side-channel attacks are attacks on computing systems that utilize unconventional interfaces exposed by the target system. An attacker can use these interfaces to extract data from the system, as well as modify the system's behavior.

The most common examples of these interfaces are the physical aspects of the system implementation, such as execution time, electromagnetic radiation, and power consumption.
For example, by measuring the exact duration of a cryptographic computation, an attacker can deduce information about the secret key used in the computation~\cite{Kocher96,Brumley11}.
In another example, electromagnetic radiation was applied to a tamper-resistant device to trigger faults in its computation, modifying its behavior and revealing information on the device's secret key~\cite{Boneh97,Biham97}.

In fact, side-channel attacks that use physical imperfections are so common that some consider an attack to be a side-channel attack \textit{if and only if} it is based on the physical implementation of the device \cite{NIST-SP-800-63-3}.

\subsection{Quantum Side-Channel Attacks and Quantum State-Channel Attacks}
As part of our analysis of QKD systems through cybersecurity concepts and methodologies, we can apply the notion of side-channel attacks to QKD systems.

Previous works have used intuition from classical cybersecurity and referred to all attacks on QKD systems that utilize physical imperfections as ``side-channel attacks'' (see, e.g.,~\cite{Nauerth2009}).
However, there is a clear issue with this approach: it classifies almost every attack as a ``side-channel attack'', rendering the definition useless.
In QKD implementations, the device that prepares the signals and the device that measures them perform a physical process, and the signals themselves are physical, photonic pulses. 
Thus, if the criterion for a side-channel attack is to utilize physical imperfections, any attack that utilizes an imperfection in the preparation device, the measurement device, or the signals is a side-channel attack.

We resolve this issue by changing the focus on what defines an interface as a side channel: instead of focusing on the physics of the implementation, we focus on what interfaces should (and should not) be exposed in a device.
Once we understand what interfaces a QKD device intends to expose, we can understand what the unconventional interfaces of the device are, and, respectively, what should be considered a side channel.

In a transmitting device (Alice), the conventional interface is the device's transmissions: the space of states that the device can transmit.
In a measuring device (Bob), the conventional interface is the device's measurement of incoming signals, which is modeled as the space of states that affect Bob's measurements.
When Eve performs an attack, the attack's input includes the states that arrive from Alice's device, the implementation's physical environment, and Eve's private ancillas, while the attack's output includes the states it sends to Bob's device, the physical environment, and Eve's private ancillas.

Thus, we view side-channel attacks on QKD as attacks that depend on more than Alice's sent states, or affect something other than the space that affects Bob's measurements (or the ancillas Eve can save inside her system).
We arrive at the following definition:

\begin{definition}
\label{def:side_channel}
    An attack is a Quantum Side-Channel Attack if either of the following two conditions is true:
    \begin{enumerate}
        \item its input non-trivially depends on more than the states Alice sends to Bob,
        \item or its output non-trivially affects more than the space that affects Bob's measurements and the attacker's private ancillas.
    \end{enumerate}
\end{definition}

Conversely, we define the term ``Quantum State-Channel Attacks'' to refer to attacks on QKD systems that are not Quantum Side-Channel Attacks and do not use elements outside of Alice and Bob's state spaces:

\begin{definition}
\label{def:state_channel}
    An attack is a Quantum State-Channel Attack if the following two conditions are both true:
    \begin{enumerate}
        \item its input non-trivially depends only on the states Alice sends to Bob,
        \item and its output non-trivially affects only the space that affects Bob's measurements and the attacker's private ancillas.
    \end{enumerate}
\end{definition}

Note that Quantum Side-Channel Attacks can utilize resources that cannot be used in Quantum State-Channel Attacks.
However, those resources are limited by standard working assumptions on the implementation, as described in international standards such as~\cite{iso-qkd-standard-pt-1}. Specifically, when considering attacks in this paper, we assume that each QKD device operates inside a shielded environment (while the channels connecting them are not shielded); that the users of the legitimate devices (Alice and Bob) are trusted; and that the pre-shared secrets of the devices are secure.
Under these assumptions, the definitions of Quantum Side-Channel Attacks and Quantum State-Channel Attacks are complementary.
However, there can exist attacks that violate these assumptions: for example, an attack scenario where Eve installs a camera in Bob's lab and sees him entering his basis choices.

Also note that, as an edge case, it is possible for Eve to construct an attack that only relies on Alice's space and only affects Bob's space, but also releases some other side-effect information that does not affect Alice or Bob. Without the side-effect, this attack would be considered a Quantum State-Channel Attack, but according to our definition, the added side-effect technically makes the full attack a Quantum Side-Channel Attack. We thus consider this edge case as a ``trivial'' case of Quantum Side-Channel Attacks.

Further note that Definitions~\ref{def:side_channel} and~\ref{def:state_channel} are somewhat informal: a rigorous formulation would require a precise mathematical definition of Alice's and Bob's enlarged Hilbert spaces corresponding to the physical reality where \textit{both} Alice and Bob could be imperfect, in addition to Eve's most general attack on them. The precise definition of those spaces and their relations is complex and subtle (see, e.g.,~\cite{gelles_qsa_2007} for such an attempt) and is left for future work.

There have been past attempts to classify attacks that do not rely on the conventional interfaces of the QKD devices.
Vakhitov, Makarov, and Hjelme~\cite{vakhitov_large_2001} used the term ``conventional optical eavesdropping'' to describe attacks where \textit{``Eve can get information by using loopholes in Alice’s and Bob’s optical set-up rather than by measuring the transmitted quantum states''}.
This term is, in fact, a special case of our definition.
While forcing leakage from the optical setups of Alice and Bob certainly qualifies as a side channel, our definition does not require that we not measure the transmitted quantum states.
In fact, as shown below, several Quantum Side-Channel Attacks collect data from QKD devices through side channels, or manipulate QKD devices through side channels, in order to perform more successful measurements of the transmitted quantum states.
Furthermore, there are other elements besides the optical setup that can be used to launch a Quantum Side-Channel Attack, as explored below.

\subsection{Examples}
\label{subsec:side_channel_examples}
We will now analyze several examples of attacks on QKD implementations under our new definitions.

\subsubsection{Quantum Side-Channel Attacks}

\paragraph*{The Large Pulse attack~\cite{vakhitov_large_2001}:}
This attack works against QKD devices where the legitimate party's private choices (specific state for Alice, measurement basis for Bob) affect the device's optical configuration.
In this attack, Eve sends a high-intensity pulse to the device and uses the reflected light to determine the private choice.
If Eve learns the used basis (from either Alice or Bob), she can use this data to perform a perfect measure-resend attack. If she learns Alice's chosen state, no further action is required.
Since this attack depends on more than Alice's sent states (specifically, the device's optical configuration), it is a Quantum Side-Channel Attack.

\paragraph*{The Injection-Locking attack~\cite{Pang2020}:}
In this attack, Eve uses specialized pulses that enter Alice's laser, forcing each of Alice's pulses to have a different wavelength depending on the encoded qubit state.
This allows Eve to measure the frequency of each transmission and learn Alice's bit perfectly.
Since this attack affects Alice's device operation, it is a Quantum Side-Channel Attack.

\paragraph*{The Bright Illumination attack~\cite{lydersen_blinding_commercial_2010}:}
We analyze the Bright Illumination attack in Section~\ref{sec:bi_analysis} and show it to be a Quantum Side-Channel Attack.

\subsubsection{Quantum State-Channel Attacks}

\paragraph*{The photon-number splitting attack~\cite{PNS}:}
This attack, discussed in Section \ref{subsubsec:qkd_vulnerabilities}, utilizes a vulnerability in Alice's transmitter, which makes Alice's sent states sometimes include more than one photon.
Eve checks the number of photons in a pulse, and if there is more than one, she saves it as her private ancilla and sends the rest to Bob. 
Eve measures her saved photon after the bases are published, gaining perfect knowledge of the secret bit.
Since the attack only utilizes Alice's signal and generates states that Bob can measure, it is a Quantum State-Channel Attack.

\paragraph*{The Time-Shift attack~\cite{qi_time-shift_2006}:}
This attack works against receivers with different detectors for different bit values, where the detection sensitivities of each detector are unequal.
Eve selectively delays Alice's signal, independently of Alice's choice of state, to make one bit value much more likely in Bob's detection.
The input of the attack is exactly Alice's signal, and the output of the attack is time-shifted states, inside the space of states that Bob's device can measure.
As such, the Time-Shift attack is a Quantum State-Channel Attack.

\paragraph*{Reversed-Space Attack example from Section~\ref{subsec:RSA-example}:}
This attack was deliberately constructed to affect all of (and only) the extended Hilbert space that affects Bob's measurements, and it does not depend on any input other than Alice's signal.
As such, this attack too is a Quantum State-Channel Attack.

\paragraph*{Trojan-Pony Attack~\cite{gottesman_security_2004}:}
In this attack, Bob's vulnerability is that his detectors cannot count the number of photons that arrive in a pulse, and he interprets all ``double-click'' events (where more than one detector clicks) as signal loss, and not as an invalid result.
Eve can use this vulnerability to launch a faked-states attack, sending states that Bob will either measure as a loss or as Eve's choice of state.
Since the attack assumes Alice is ideal and only affects the space Bob measures, it is a Quantum State-Channel Attack.

\paragraph*{Imperfect Faraday Mirror Attack~\cite{wang_faraday_2013}:}
In this attack, a fault in Alice's optical setup makes her states span a three-dimensional Hilbert space instead of the standard two-dimensional qubit space.
Eve performs a measure-resend attack, achieving a stronger distinction between the four states in the three-dimensional space, gaining significant information on the key while keeping the error rate below the required threshold.
Since the attack relies on Alice's sent states (and outputs valid qubits), it is also a Quantum State-Channel Attack.

\medskip

Further classification of attacks under the Quantum Side-Channel Attack and Quantum State-Channel Attack definitions can be found in Section~\ref{sec:applicability}.

\subsection{Application to Security Analyses}
As shown in the above subsection, a wide variety of side channels can be used to attack QKD implementations.
For an implementation to be proven secure, all these side channels must be considered in the theoretical model of the implementation, though we are not aware of security proofs that use this full modeling.
Alternatively, a separate engineering analysis can examine the physical QKD device and its potential side channels and ensure that these side channels cannot be exploited by Eve.

\section{Putting It All Together: Analysis of the Bright Illumination Attack}
\label{sec:bi_analysis}

In this section, we analyze the ``Bright Illumination'' attack and its inner workings using the novel tools we presented in the previous sections.
We show that a structured security analysis could have found the Bright Illumination attack, using the tools we propose in this paper and based on an understanding of the importance of practical vulnerability research.

\subsection{The Bright Illumination Attack}
\label{subsec:bi_explanation}
The ``Bright Illumination'' attack~\cite{makarov_blinding_first_2009, lydersen_blinding_commercial_2010,lydersen_blinding_thermal_2010, sague_blinding_pulsed_2011}
is an implementation attack that changes the behavior of QKD receivers by exposing them to high-intensity light.
This attack was implemented against a variety of QKD implementations, including implementations from recent years~\cite{Gras2020,Gao2022}, using various photodetectors and implementing various protocols~\cite{lydersen_blinding_snspd_2011, chaiwongkhot_blinding_tes_2022}.
In this section we analyze the specific attack and implementation shown in~\cite{sague_blinding_pulsed_2011};
however, our analysis can be applied with minor modifications to other Bright Illumination attacks against various QKD implementations.

\begin{figure}[ht]
    \centering
    \includegraphics[width=\linewidth]{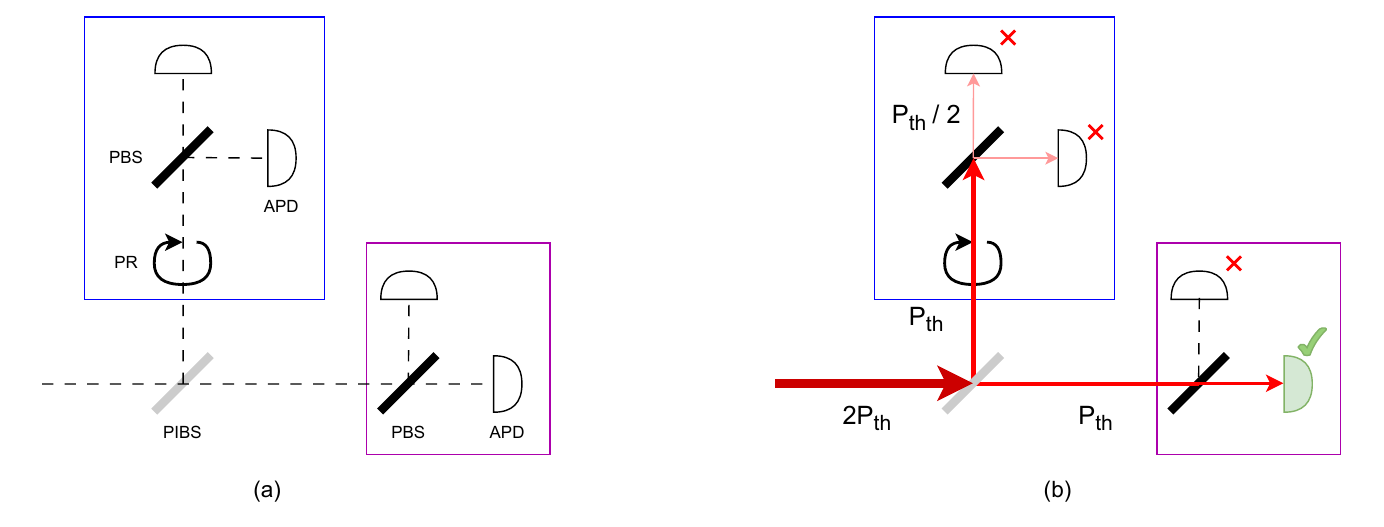}
    \caption{(a) Structure of the QKD receiver targeted in the Bright Illumination attack. PIBS: polarization-independent beam splitter. PBS: polarizing beam splitter. PR: polarization rotator. APD: avalanche photodiode. (b) Propagation of Eve's high-intensity pulse through the QKD receiver.}
    \label{fig:bi-setup}
    \Description[(a) shows basis choice by PIBS, and one PBS and two APDs for each basis measurement. In (b), Eve activates only one detector.]{Subfigure (a) shows a QKD receiver made of a PIBS connected to a PBS on its right and a PR followed by a PBS above. Each PBS is connected to two APDs, one on its right and one above. Subfigure (b) shows the flow of light through the device in a Bright Illumination attack. A wide red arrow, representing light of intensity "2 P threshold", enters the PIBS. Two regular-width arrows, representing light of intensity "P threshold", then connect the PIBS to each one of the PBSes (the PBS on the right and the PBS above, respectively). For the PBS on the right, the entire arrow (light of intensity "P threshold") continues to the right APD, while the APD above does not receive any light, so a check mark appears near the right APD, and an X mark appears near the APD above. In contrast, for the PBS above, the PBS is connected to each APD with a thin-width arrow (each arrow representing light of intensity "P threshold / 2"), and an X mark appears near each one of those APDs.}
\end{figure}

\paragraph{Receiver Structure}
In the QKD system targeted by~\cite{sague_blinding_pulsed_2011}, the implemented protocol is polarization-based BB84, and measurement is performed using a passive basis choice and four detectors, as shown in Figure~\ref{fig:bi-setup}~(a).

The passive basis choice in the receiver is implemented using a polarization-independent beam splitter, 
which leads to two sub-components. Each sub-component measures in a different basis: the computational basis measurement (marked in purple) is performed using a polarizing beam splitter and two detectors, one in each output arm.
The Hadamard basis measurement (marked in blue) is performed in the same way, with the addition of a polarization rotator before the beam splitter
to transform the Hadamard basis states into computational basis states.

The detectors used in this implementation are Avalanche Photodiodes (APDs).

APDs have two possible modes of operation, called the ``Geiger mode'' and ``linear mode''~\cite{makarov_blinding_first_2009}.
In Geiger mode, an APD operates as a single-photon detector: it generates a ``click'' (a macroscopic event) when it receives one or more photons. 
However, in linear mode, the APD ``clicks'' only if the incoming pulse's intensity is above a certain threshold:
if the pulse intensity is below that threshold, no effect is registered in the APD (there is no click). 
Thus, an APD in linear mode is ``blind'' to low-intensity pulses, including single photons.

In QKD setups, APDs operate in Geiger mode as single-photon detectors.
However, an APD can be forcibly reconfigured to linear mode if it is bombarded with high-intensity light (or, potentially, by heating the APD~\cite{lydersen_blinding_thermal_2010}).

\paragraph{Attack Structure}
The Bright Illumination attack (shown in~\cite{sague_blinding_pulsed_2011}) relies on APDs having an undesired, linear mode of operation, and the ability to force an APD to operate in the linear mode by sending high-intensity light to it.

In the attack, Eve sends high-intensity light pulses to Bob's device \textit{before} the intended detection window.
This makes the APDs operate in linear mode during the detection window, making them blind to low-intensity pulses.
Once the APDs are in linear mode, Eve performs the ``measure-resend'' attack described in Figure~\ref{fig:bi-setup}~(b):
she picks a specific measurement basis, measures Alice's signal in that basis, and then generates a signal in the same polarization whose intensity is slightly above $2P_\mathrm{th}$,  where $P_\mathrm{th}$ is defined as the minimal intensity the APDs can detect in linear mode.

When the pulse enters the polarization-independent beam splitter in Bob's device, it is split into two pulses with intensity slightly higher than $P_\mathrm{th}$.
In the sub-component that measures in the basis Eve chose, the polarizing beam splitter will transfer the entire pulse to one arm, which will cause a click in the intended detector.
In the other sub-component, the polarizing beam splitter will split the pulse into two parts with intensity around $P_\mathrm{th}/2$, which will not cause a click in either detector.

Thus, Eve's attack results in a guaranteed click in a detector of her choosing, causing Eve and Bob to receive the same measurement results.
Using this information, Eve can mimic Bob's classical process by listening to the classical channel. This leaves Eve with full knowledge of the final key when the protocol concludes.

\subsection{Is Bright Illumination a Quantum Side-Channel Attack?}
\label{subsec:bi_side}
As discussed in Sections~\ref{subsec:bi_explanation} and~\ref{subsec:bi_fuzz}, the Bright Illumination attack depends on two separate stages:
the first stage is to modify the behavior of Bob's device (causing the APDs to switch to the linear mode), and the second stage is a measure-resend attack on the modified device.
Since in the first stage of the attack, Eve sends blinding pulses that change the behavior of the device (and are not simply measured), the Bright Illumination Attack is a Quantum Side-Channel Attack according to Definition~\ref{def:side_channel}.

\subsection{Finding the Vulnerabilities via Quantum Fuzzing}
\label{subsec:bi_fuzz}
In this subsection, we show that applying Quantum Fuzzing to the QKD implementation shown above can reveal the vulnerabilities enabling the Bright Illumination attack.

When using the Quantum Fuzzing strategy defined in Section~\ref{subsec:fuzz_strategy}, 
the procedure begins by testing valid protocol states and then gradually modifies certain elements in the states.
We start by performing a small set of modifications, and if the results do not expose interesting behavior, we continue by testing either a broader range of values or additional values inside the existing range.
Since our fuzzing strategy modifies the intensity of pulses relatively early in the procedure,
the following property will be revealed after a few test cases:

\begin{property}
\label{prop:bi-basic}
    \textbf{Blinding.} Sending a high-intensity pulse to Bob's device causes a loss, instead of burning the device or showing an invalid result (such as a click in two detectors at the same time).
\end{property}

As described in Section~\ref{subsec:bi_explanation}, this effect is caused by the bright pulse forcing Bob's APDs into the linear mode.
This result is highly unexpected: as the intensity of the incoming pulse increases, one should expect the chance of pulse detection to be higher --- but our high-intensity pulse does not cause a detection event at all.

If the Quantum Fuzzing process were to stop after this observation, the useful nature of the linear mode would not be revealed. However, as discussed in Section~\ref{subsec:fuzz_strategy}, if a test case triggers an ``interesting'' effect on a device, our strategy suggests combining it with other test cases, in the hope of expanding the effect.
Employing this strategy, we would send input states that include a high-intensity pulse, followed by another photon pulse. When we test various intensity values for the second pulse, gradually increasing the intensity range and granularity, we reveal two more properties, one after the other:

\begin{property}
    \textbf{Weak Pulses under Blinding.} Sending a high-intensity pulse to Bob's device, followed by another pulse with a small photon number, will not cause a detection event.
\end{property}

\begin{property}
\label{prop:bi-full}
    \textbf{Strong Pulses under Blinding.} Sending a high-intensity pulse to Bob's device, followed by another pulse with a high photon number, will \emph{always} cause a detection event in the detector matching the pulse's polarization, and not in detectors used for the other basis.
\end{property}

These two properties reveal how Bob's device acts when it is ``blinded'' (that is, when the APDs in Bob's device are in the linear mode).
In fact, these properties reveal Measurement Space Vulnerabilities and Interpretation Vulnerabilities in Bob's device (see Definitions~\ref{def:msv} and~\ref{def:iv}): when Bob's device is blinded, it measures four states (the high-photon-number variants of the BB84 states) that it should not measure, and interprets them as valid if they match his basis choice, or as a loss if they do not match his basis choice.

Now that we have revealed the Measurement Space Vulnerabilities and Interpretation Vulnerabilities in the device, we can build an exploit for them using the Reversed-Space method.

\subsection{Applying Reversed-Space}
\label{subsec:bi_rsa}
In the previous subsection, we used Quantum Fuzzing to find 
a set of Measurement Space Vulnerabilities and Interpretation Vulnerabilities.
We now wish to build a Reversed-Space Attack using these vulnerabilities, according to the procedure described in Section~\ref{sec:RSA}.
For the sake of readability, the computation can be found in Appendix~\ref{app:bi_rsa_exploit}.
The result of the computation is an attack of the form:

\begin{equation}
	\begin{array}{rcccccl}
		U_{\sE} \ket{0}_{\sE}\ket{0}_{\sA} &=&
        p\ket{E_0}\ket{0}^\mathrm{bright} &+&
        q\ket{E_2}\ket{+}^\mathrm{bright} &+&
        q\ket{E_3}\ket{-}^\mathrm{bright}, \\
            U_{\sE} \ket{0}_{\sE}\ket{1}_{\sA} &=&
        p\ket{E_1}\ket{1}^\mathrm{bright} &+&
        q\ket{E_2}\ket{+}^\mathrm{bright} &-&
        q\ket{E_3}\ket{-}^\mathrm{bright}, \\
        
	\end{array}
\end{equation}
where $p$ and $q$ are non-negative real numbers that satisfy
\begin{equation}
    p^2 + 2 q^2 = 1
\end{equation}
and the state $\ket{\psi}^\mathrm{bright}$ is a high-photon-number variant of the single-photon state $\ket{\psi}$.

To give some intuition to the result, choosing $p = 1,\; q = 0$ gives a CNOT attack from Alice's computational states
to the ``blinding-computational'' states.
Choosing $p = 0,\; q = \frac{1}{\sqrt{2}}$ gives a CNOT attack from Alice's Hadamard states to the ``blinding-Hadamard'' states. Intermediate attacks with $0 < p < 1 \; , \; 0 < q < \frac{1}{\sqrt{2}}$ resemble the original Bright Illumination attack, which sometimes sends high-photon-number computational basis states and sometimes sends high-photon-number Hadamard basis states.

Similarly to Section~\ref{subsec:RSA-example}, this attack breaks the security (and the robustness: see~\cite{BKM2007}) of the implementation, because Eve gains \textit{full} information on the key without inducing any error and without causing the protocol to abort.

\medskip

The computation of this attack completes the proof that Bright Illumination attacks can be constructed as Reversed-Space Attacks, using vulnerabilities found via Quantum Fuzzing: a system designer with black-box access to the QKD system and no knowledge of its internal workings could have constructed the full attack.

\section{Applicability to Current Implementation Attacks}
\label{sec:applicability}
In this section, we explore the relationship between different QKD attacks and the novel methodologies and concepts defined in this paper. Our goal is to provide a broader context for our results and emphasize their usefulness in classifying existing attacks.

\begin{figure}[htb]
	\centering
	\includegraphics[width=\linewidth]{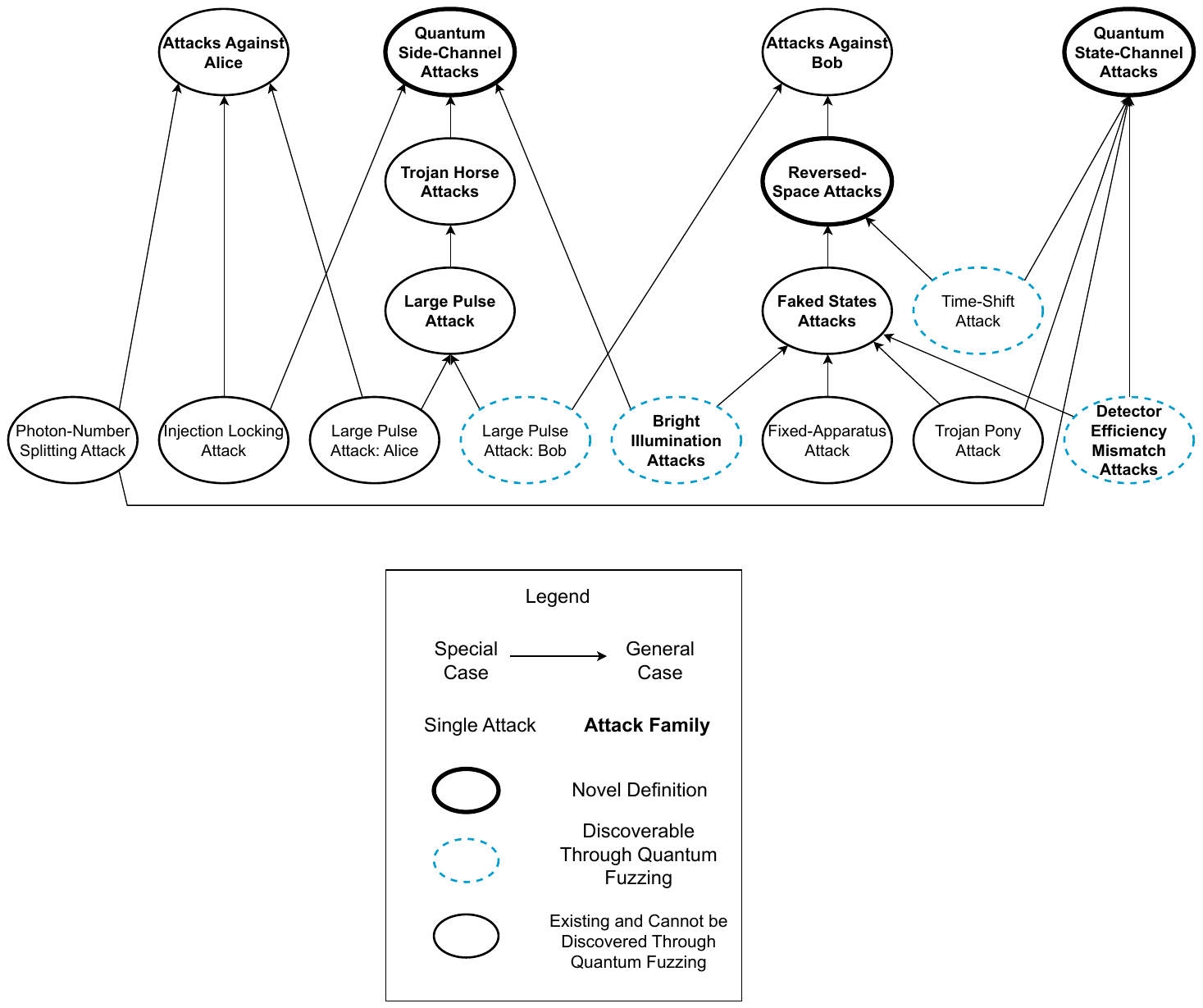}
		\caption{A representation of the relations between QKD attacks and our results.}
        \Description[Special cases of attacks are displayed at the bottom of the diagram, and are classified into more general classes at the top.]{The bottom of the diagram consists of nine attacks: Photon-Number Splitting Attack, Injection Locking Attack, Large Pulse Attack: Alice, Large Pulse Attack: Bob, Bright Illumination Attacks, Fixed-Apparatus Attack, Trojan Pony Attack, Time-Shift Attack, and Detector Efficiency Mismatch Attacks. Large Pulse Attacks are a general case of Large Pulse Attack: Alice and Large Pulse Attack: Bob. Trojan Horse Attacks are a general case of Large Pulse Attacks.
        Faked States Attacks are a general case of Bright Illumination Attacks, Fixed-Apparatus Attack, Trojan Pony Attack, and Detector Efficiency Mismatch Attacks. Reversed-Space Attacks are a general case of Faked States Attacks and Time-Shift Attack. Quantum Side-Channel Attacks are a general case of Injection Locking Attack, Trojan Horse Attacks, and Bright Illumination Attacks. Quantum State-Channel Attacks are a general case of Photon-Number Splitting Attack, Time-Shift Attack, Trojan Pony Attack, and Detector Efficiency Mismatch Attacks. Attacks Against Alice are a general case of Photon-Number Splitting Attack, Injection Locking Attack, and Large Pulse Attack: Alice. Attacks Against Bob are a general case of Large Pulse Attack: Bob and Reversed-Space Attacks. The attacks in the diagram that are novel definitions in the paper are Reversed-Space Attacks, Quantum Side-Channel Attacks, and Quantum State-Channel Attacks.
        The attacks in the diagram that are discoverable through Quantum Fuzzing are Large Pulse Attack: Bob, Bright Illumination Attacks, Time-Shift Attack, and Detector Efficiency Mismatch Attacks.
        The Attack Families in the diagram are Bright Illumination Attacks, Detector Efficiency Mismatch Attacks, Large Pulse Attacks, Faked States Attacks, Trojan Horse Attacks, Reversed-Space Attacks, Attacks Against Alice, Attacks Against Bob, Quantum Side-Channel Attacks, and Quantum State-Channel Attacks.}
\label{fig:attack_relations}
\end{figure}

To illustrate the relationships between different attacks on QKD implementations and our methodologies, we present a visual diagram in Figure~\ref{fig:attack_relations}, accompanied by an explanation of the results and insights presented in the diagram.
High-level details on each attack in the diagram can be found in Appendix~\ref{app:attack_list}.

The diagram is constructed as a graph, where each node represents an attack (or attack family), and an edge from one node to another means that the former is a special case of the latter.

We will now discuss the relations between existing QKD attacks and each of our results separately.

\subsection{Relationship to Quantum Side-Channel Attacks}
\begin{observation}
There are Quantum Side-Channel Attacks and Quantum State-Channel Attacks against Alice and against Bob.
\end{observation}

The diagram, as well as Section~\ref{subsec:side_channel_examples}, shows that Quantum Side-Channel Attacks are not only a theoretical definition: there are several practical examples of such attacks both against Alice and against Bob, including the Large Pulse attacks, the Injection Locking attack, and the Bright Illumination attack family.

However, in contrast to earlier definitions of side-channel attacks in QKD (see, e.g.,~\cite{Nauerth2009}), many practical attacks are \textit{not} Quantum Side-Channel Attacks: many (if not most) well-known attack examples do not use elements that are external to Alice and Bob's realistic protocol, and are therefore not Quantum Side-Channel Attacks, but rather Quantum State-Channel Attacks. Examples of such attacks are the Photon-Number-Splitting attack, the Trojan Pony attack, and the Time-Shift attack.
Hence, our definitions of Quantum Side-Channel Attacks and Quantum State-Channel Attacks represent a meaningful distinction between implementation attacks.

\subsection{Relationship to Reversed-Space Attacks}
\begin{observation}
    Reversed-Space Attacks are a strict generalization of Faked-States attacks, and include some Quantum Side-Channel Attacks, but do not include all attacks against Bob.
\end{observation}

Reversed-Space Attacks are a generalization of Faked-States attacks for the following reason: since Faked-States attacks rely on the existence of ``faked states'' that force a specific interpretation in Bob's measurement device, they rely on Measurement Space Vulnerabilities and Interpretation Vulnerabilities.
Hence, all Faked-States attacks are in particular Reversed-Space Attacks.
However, the opposite is not true: the Time-Shift attack is a Reversed-Space Attack which is not a Faked-States attack, because it depends on an enlarged measured space, but it does not measure Alice's signal as is required in Faked-States attacks~\cite{makarov_faked_2005}.

The Bright Illumination attack shown in Section~\ref{sec:bi_analysis} is an example of a Reversed-Space Attack that is also a Quantum Side-Channel Attack.

Finally, the Large Pulse attack (against Bob) is an example of a non-Reversed-Space Attack against Bob, since it does not depend on an enlarged measurement space in Bob's interpretation.

\subsection{Relationship to Quantum Fuzzing}
\begin{observation}
    Quantum Fuzzing can discover both Quantum Side-Channel Attacks and Quantum State-Channel Attacks, but cannot discover all attacks against Bob. Furthermore, our current strategy cannot be applied against Alice's device.
\end{observation}

Since Quantum Fuzzing tests the behavior of a device by sending different input states to it, it can reveal the vulnerabilities behind several attacks.
Four attacks that are triggered by sending specific input states to Bob's device are the Large Pulse attack (against Bob), the Bright Illumination attacks, the Detector Efficiency Mismatch attacks, and the Time-Shift attack.
The first two attacks are examples of Quantum Side-Channel Attacks, and the last two attacks are Quantum State-Channel Attacks.

However, not all attacks against Bob can be found via Quantum Fuzzing, since some attacks rely on elements that are outside of the device's input channel.
For example, the Fixed-Apparatus attack that uses the blocked arm inside Bob's device, which is not available through the input channel.

Additionally, since our Quantum Fuzzing strategy relies on the existence of valid protocol states as initial test cases, and Alice's device is not supposed to receive any outside input, our strategy cannot be applied to discover attacks against Alice.
However, a different Quantum Fuzzing strategy can potentially reveal attacks on Alice's device that require external input from Eve, such as the Injection Locking attack and the Large Pulse attack (against Alice).

\section{Conclusions and Future Work}
\label{sec:conclusions}
Our work defines a quantum cybersecurity approach for systematically analyzing the security of QKD implementations, inspired by cybersecurity research on classical computing systems.
We have shown a fundamental connection between imperfections and attacks on QKD implementations, and the classical notions of vulnerabilities, attack surfaces, and exploits (Section~\ref{sec:vulns_exploits}).
We have also defined common vulnerability types in QKD devices, shown the role of attack surfaces in the feasibility of attacks on QKD devices, and examined generic exploit methods for implementation attacks.

Expanding on the connection between classical cybersecurity and QKD implementations, we have presented three additional contributions.
First, through our definition of Quantum Fuzzing (Section~\ref{sec:fuzzing}), which is the first vulnerability research method for QKD implementations, QKD system designers can search for issues in their implementation that they did not know existed.
Second, our Reversed-Space Attacks methodology (Section~\ref{sec:RSA}) is a new generic exploit which computes and utilizes the QKD receiver's attack surface, allowing system designers and attackers to build more sophisticated implementation attacks.
Third, our definitions of ``Quantum Side-Channel Attacks'' and ``Quantum State-Channel Attacks'' (Section~\ref{sec:side_channels}) account for the special nature of side channels in QKD devices, and emphasize that the security of QKD devices can be compromised by elements both inside and outside of Alice's transmission and Bob's measurement.

The paper concludes with two additional applications of our newly-defined concepts and methodologies.
First, we have revisited the Bright Illumination attack (Section~\ref{sec:bi_analysis}) and demonstrated how our tools could have predicted this attack through a rigorous security analysis.
Second, we have classified existing attacks using our new definitions (Section~\ref{sec:applicability}), emphasizing the applicability of our tools to a large number of attacks and attack families.

The goal of our work is to create a connection between classical and quantum implementation attacks and allow greater collaborations between classical cybersecurity experts and QKD experts.
Quantum researchers can borrow approaches and results from classical cybersecurity in order to define new attacks and defend against them.
Cybersecurity researchers can understand attacks on QKD implementations through their own perspective and apply their personal experience and expertise.
This, in turn, can greatly improve the practical security of experimental and commercial QKD devices.

By building new tools for analyzing threats from practical adversaries, we aim to bridge the gap between practical QKD security research that only deals with all-knowing, all-powerful adversaries and classical cybersecurity that often deals with realistic, limited adversaries and what they can achieve.
The benefits of this added perspective are clear: for instance, awareness of the need to do vulnerability research, together with the tools for exploiting space-enlargement vulnerabilities, could have revealed the Bright Illumination attack and enabled QKD system designers to discover it before potential adversaries.
More importantly, these tools can help discover further attacks in the future.

Since our work focuses on attacks on imperfect QKD devices, one may try to partially solve the problem by using QKD protocols with an untrusted center party (typically named Charlie): since an untrusted party is assumed to be malicious, imperfections in its implementation cannot harm security.
However, even in such protocols, imperfections in the trusted parties can still harm security, and our tools can still be used to analyze them.
While entanglement-based protocols (such as BBM92~\cite{BBM92}) protect against state generation imperfections, attacks against the measurement device (such as Reversed-Space Attacks) are still possible.
Similarly, in protocols with measurement by an untrusted center (such as BHM96~\cite{BHM96} or MDI-QKD~\cite{Xu2020}), attacks against the state preparation devices (such as the Photon-Number-Splitting and Large Pulse attacks~\cite{PNS,vakhitov_large_2001}) are still possible.
While Device-Independent QKD protocols~\cite{di_qkd} assume \textit{all} devices to be untrusted and thus theoretically secure against implementation attacks, they still depend on many assumptions, including the assumption that Alice and Bob's input choices are random and independent.
Failure to satisfy these assumptions in practical implementations can, in fact, compromise the security of the protocol, as seen, for example, in~\cite{Gerhardt2011}.

Several directions we consider promising for future research include the application of our results to commercial QKD implementations;
the development of physical Quantum Fuzzing devices, as well as additional strategies for Quantum Fuzzing (especially against QKD state preparation devices);
and the continued exploration of vulnerability types in QKD devices and generic exploit methods, in order to provide a more complete map of QKD implementation attacks.

To sum up, in this work we have begun to bridge the gap between practical security analyses of QKD implementations and the decades-long extensive research in the field of classical cybersecurity.
We hope that this research improves the practical security of future QKD products and enhances their usefulness in real-world systems.

\section*{Acknowledgements}
The work of R.L.\ was partly supported by the Government of Spain (Severo Ochoa CEX2019-000910-S, FUNQIP, and European Union NextGenerationEU PRTR-C17.I1), Fundaci\'o Cellex, Fundaci\'o Mir-Puig, Generalitat de Catalunya (CERCA program), and the European Union (Quantum Secure Networks Partnership (QSNP), Grant 101114043).
The work of T.M.\ was partly supported by the Technion’s Helen Diller Quantum Center (Haifa, Israel).

\bibliographystyle{unsrturl}
\bibliography{joint_paper_refs.bib}

\appendix

\section{Application of Linear Optical Devices on Photonic States}
\label{app:quantum_optics}

In this appendix, we discuss physical implementations of QKD schemes using photons.
We describe the unitary transformations that characterize various optical devices commonly used in contemporary QKD implementations, such as beam splitters, phase shifters, and interferometers.
We explain the quantum state of a photon (or more generally, a pulse of photons) passing through these optical devices.

\subsection*{Symmetric Beam Splitter}
\begin{figure}[htb] 
 \centering 
\includegraphics{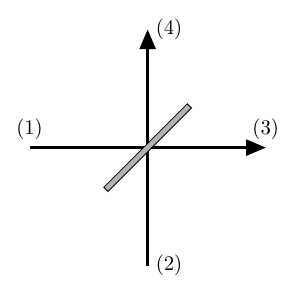}

\caption{A symmetric beam splitter with two input modes, (1) and (2), and two output modes, (3) and (4). 
} 
\Description[The Beam Splitter is placed diagonally with numbered inputs and outputs.]{The edges of the Beam Splitter are at the top-right corner and the bottom-left corner. Input mode (1) is on the left, input mode (2) is at the bottom, output mode (3) is on the right, and output mode (4) is at the top.}
 \label{fig:BS} 
\end{figure}

A beam splitter has two input arms (modes 1 and 2) and two output arms (modes 3 and 4),
as depicted in Figure~\ref{fig:BS}. 
Assuming that the beam splitter is symmetric,
each entering photon has equal amplitudes for transmittance and reflection;
the transmitted part keeps the same phase as 
the incoming photon, while the reflected
part gets an extra phase of $e^{i\pi/2} \equiv i$. Specifically, $ \fet{0{,}1}_{2,1} \to
 \frac{1}{\sqrt{2}}(\fet{0{,}1}_{4,3} + i\fet{1{,}0}_{4,3})$ and
$\fet{1{,}0}_{2,1} \to  \frac{1}{\sqrt{2}}(i\fet{0{,}1}_{4,3} + \fet{1{,}0}_{4,3})$.
Thus, for a single photon state in a general superposition between the two input arms,
the transformation is given by
\begin{equation}\label{eqn:bs}
\alpha\fet{0{,}1}_{2,1}+ \beta\fet{1{,}0}_{2,1}
\mapsto \frac{\alpha+i\beta}{\sqrt{2}}\fet{0{,}1}_{4,3}+ 
\frac{i\alpha+\beta}{\sqrt{2}}\fet{1{,}0}_{4,3}.
\end{equation}

It is important to note that when a single photon
in a single mode enters a beam splitter
from one arm, and nothing --- that is, the vacuum state
--- enters the other arm (say, $\alpha=1; \beta=0$), 
there are still two input modes and two output modes. 
This means that the other (vacuum)
entry must be considered as an additional mode: an ancilla carrying
no photons.

\subsubsection*{Action on a High-Photon-Number Pulse}
As discussed in~\cite{Agarwal2012}, the result of a symmetric beam splitter on a pulse with $n$ photons,
denoted as $\fet{0,n}_{2,1}$, is given by:
\begin{equation}
	\ket{\psi_\mathrm{out}} = \frac{1}{\sqrt{2^n}}\sum_{k=0}^n \sqrt{\binom{n}{k}}\fet{k,n-k}_{4,3}.
\end{equation}

\subsection*{Phase Shifter}
A controlled phase shifter $P_\phi$  performs a phase 
shift on the input state by a given phase $\phi$ ---
that is, $P_\phi(\fet{n}) = e^{i n \phi}\fet{n}$; see~\cite{QOpticsBook} for more details.

\subsection*{Mach-Zehnder Interferometer}
\begin{figure}[htb] 
 \centering 
\includegraphics{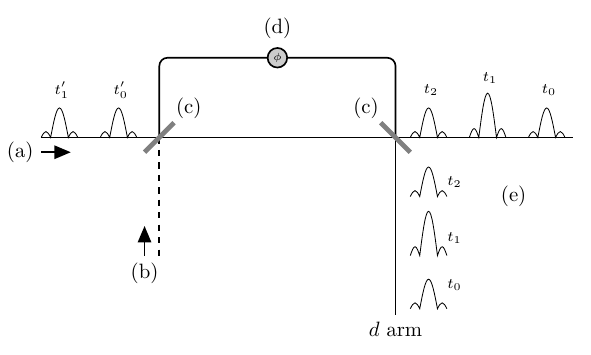}
 \caption
 {A Mach-Zehnder interferometer. 
 (a) An input qubit. The time-difference between the two incoming modes is 
identical to the difference between the two arms; 
(b) a vacuum state entering the second (blocked) arm;
 (c) beam splitters; (d) phase shifter $P_\phi$; 
 (e) six output modes.  
 }
 \Description[Two possible input pulses, two arms (long and short), and six possible output pulses on the interferometer's paths.]{Two pulses denoted "t prime 0" and "t prime 1" are at the left input arm of the first beam splitter, denoted (a). A dashed line is at the bottom input arm of the first beam splitter, denoted (b). The short path between the two beam splitters includes nothing, and the long path between them includes a gray circle with a "phi" symbol denoting a phase shifter (denoted (d)). Both at the right output arm and at the bottom output arm of the second beam splitter, three pulses are illustrated, denoted "t 2", "t 1", and "t 0" (a total of six pulses, denoted (e)), where "t 2" is the closest to the beam splitter, "t 1" is in the middle, and "t 0" is the farthest away from the beam splitter. The pulses denoted "t 1" in both arms are slightly larger than those denoted "t 0" and "t 2".}
 \label{fig:lab-xy} 
\end{figure}

A Mach-Zehnder interferometer~(Figure~\ref{fig:lab-xy}) is a device composed of
two beam splitters (BS) with one short path, one long path, and a 
controlled phase shifter $P_\phi$, that is placed at the long arm of the interferometer. 

We will now describe the operation of the interferometer on input states similar to those
used in the example protocol in Appendix~\ref{app:rsa_main_example}.

\medskip

In each transmission, a superposition of two (time) modes enters the interferometer
and result in a superposition of 6 modes (Figure~\ref{fig:lab-xy}).
The input modes are separated with a time difference of $\Delta T$ seconds: that is,
the first mode arrives at time $t_0'$, and the second at $t_1'=t_0'+\Delta T$.
The first pulse travels through the short arm 
in $T_{\rm short}$ seconds, and through the long arm in 
$T_{\rm long} =  T_{\rm short} +\Delta T$
seconds, where the time difference between the two arms is exactly the time difference $\Delta T$ 
between the two incoming modes. 
Due to traveling through both arms, the first mode yields outgoing pulses both at time 
$t_0 \equiv t'_0 + T_{\rm short}$ and at 
$t_1 \equiv t'_0 + T_{\rm long} = t'_0 + T_{\rm short} + \Delta T =  t_0 + \Delta T$.

When the second pulse enters the interferometer, it also travels through both arms.
Intuitively, 
the part of the $t'_1$ mode that travels through the short arm
interferes with the part of the $t'_0$ mode that travels through the long arm,
and the output exits the interferometer at $t_1$.
The part of the second pulse that travels through
the long arm exits the interferometer at time $t_2 = t_1 + \Delta T$.
As a result, there are six different possible modes at the two output arms, three in
each direction, with the two middle pulses determined by the interference
between the two pulses arriving into Bob's lab.

We shall now show the mathematical formulation behind these statements.

\subsection*{Evolution of a Single-Time-Bin Photon Through a Mach-Zehnder Interferometer}
\begin{figure}[htp]
\begin{framed}
\begin{tabular}{p{0.4\columnwidth} p{0.55\columnwidth}}
\begin{minipage}[c]{\linewidth}
\centering
\includegraphics{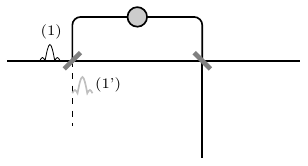}
\end{minipage}

&
\begin{minipage}[r]{\linewidth}
Pulse (1) 
		is about to enter the interferometer. 
		A vacuum ancilla (1') is added at the input 
		of the first beam splitter, $\BS_1$. 
\end{minipage}
\end{tabular}
\begin{tabular}{p{0.4\columnwidth} p{0.55\columnwidth}}
\begin{minipage}[c]{\linewidth}
\centering
\includegraphics{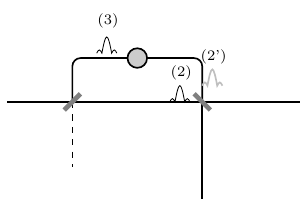}
\end{minipage}
&
\begin{minipage}[r]{\linewidth}
Pulses (1) and (1') 
		interfere in the first beam splitter ($\BS_1$)
		and yield a superposition of (2) and (3) in
		the short and long arms of the interferometer, 
		respectively: $\fet{0}_{1'}\fet{1}_1 \mapm{\BS_1} 
		(\fet{0}_{3}\fet{1}_2 + i\fet{1}_{3}\fet{0}_2)/\sqrt{2}$.
		Pulse (2) is about to enter the second beam 
		splitter ($\BS_2$), so a vacuum ancilla is added (2').   
\end{minipage}

\end{tabular}
\begin{tabular}{p{0.4\columnwidth} p{0.55\columnwidth}}
\begin{minipage}[c]{\linewidth}
\includegraphics{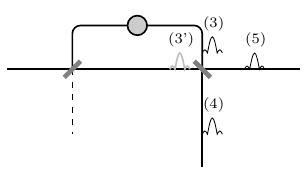}
\end{minipage}
&
\begin{minipage}[r]{\linewidth}
    Pulses (4) and (5)
		are created by pulses (2) and (2'):
		$\frac{1}{\sqrt{2}}\fet{1}_{2}\fet{0}_{2'} \mapm{\BS_2} 
		(i\fet{0}_{5}\fet{1}_4+\fet{1}_{5}\fet{0}_4)/2$.
		Pulse (3) is about to enter the second
		beam splitter, so a vacuum ancilla is added (3').
\end{minipage}

\end{tabular}
\begin{tabular}{p{0.4\columnwidth} p{0.55\columnwidth}}
\begin{minipage}[c]{\linewidth}
\centering
\includegraphics{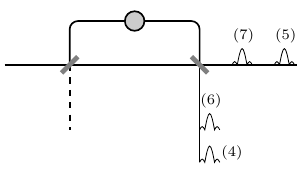}
\end{minipage}
&
\begin{minipage}[r]{\linewidth}
Pulses (6) and (7) are  
			created by the interference of (3) and (3'): 
			$\frac{i}{\sqrt{2}}\fet{0}_{3'}\fet{1}_{3}
			\mapm{\BS_2} (i\fet{0}_{7}\fet{1}_6 - 
			\fet{1}_{7}\fet{0}_6)/2$.           
\end{minipage}
\end{tabular}
\end{framed}    
     \caption{\small
		Evolution in time of a single 
		photon pulse through the interferometer with $\phi=0$: 
		$\fet{0{,}0{,}0{,}1}_{3',2',1',1} \to 
        \frac{1}{2} \left (\fet{0{,}0{,}0{,}1} - \fet{0{,}0{,}1{,}0}+ 
        i\fet{0{,}1{,}0{,}0} + i\fet{1{,}0{,}0{,}0} \right )_{6,4,7,5}$.
		The output state is denoted by modes 
		$\fet{n_{d_1}, n_{d_0}, n_{s_1}, n_{s_0}}$ that 
		correspond to modes (6), (4), (7), and (5), respectively.}
\label{fig:1evo}
\Description{fully described inside the figure.}
\end{figure}

When a single mode, carrying one or more photons,
enters the interferometer,
three ancillas in a vacuum state
are added by the interferometric setup (see Figure~\ref{fig:1evo}). 
As mentioned above, the mode that enters the interferometer at time $t'_0$, yields 
two modes at time $t_0$, and two modes at time $t_1$. 
These four output modes are: 
times $t_0$, $t_1$ at the `s' (straight) arm of the interferometer, 
and times $t_0$, $t_1$ at the `d' (down) arm of the interferometer.
A basis state of this Fock-space can be written as
$\fet{n_{d_1}, n_{d_0}, n_{s_1}, n_{s_0}}$.

Assume that a single photon enters the interferometer at
time $t'_0$. 
Using the above notations, the interferometer's transformation is given by 
\begin{equation}
\label{eqn:single-pulse-evolution}
 \fet{0{,}0{,}0}\fet{1}_{t'_0} \mapsto 
(\fet{0{,}0{,}0{,}1}-e^{i\phi}\fet{0{,}0{,}1{,}0}+i\fet{0{,}1{,}0{,}0} +i e^{i\phi}\fet{1{,}0{,}0{,}0}) \thickspace / 2 
\ .
\end{equation}
Note the three input vacuum ancillas that were added. 
Also note that a pulse sent at
a different time (say, $t'_1$ or $t'_{-1}$) 
results in the same output state, 
with appropriate delays. 
That is, a single-photon pulse entering the interferometer at time $t'_k$ 
results in the state
$(\fet{0{,}0{,}0{,}1}-e^{i\phi}\fet{0{,}0{,}1{,}0}+i\fet{0{,}1{,}0{,}0} +i e^{i\phi}\fet{1{,}0{,}0{,}0}) \thickspace / 2 $
in a Fock-space with basis states
$\fet{n_{d_{k+1}}, n_{d_k}, n_{s_{k+1}}, n_{s_k}}$.

Using the more standard notation of $\{\ket{t'_k}\}$ input states and $\{\ket{s_k}, \ket{d_k}\}$ output states
used in the rest of the paper, Eq.~\eqref{eqn:single-pulse-evolution} is given by
\begin{equation}
    \label{eqn:mzi_generic_single_photon}
	\ket{t'_k} \mapsto \left(\ket{s_k} + i\ket{d_k} -e^{i\phi}\ket{s_{k+1}} + i e^{i\phi}\ket{d_{k+1}}  \right)/2\text{ .}
\end{equation}

\subsection*{Evolution of Photon in Time-Bin Superposition Through Mach-Zehnder Interferometer}
\begin{figure}[p]
\begin{framed}
\begin{tabular}{p{0.4\columnwidth} p{0.55\columnwidth}}
\begin{minipage}[c]{\linewidth}
\centering
\includegraphics{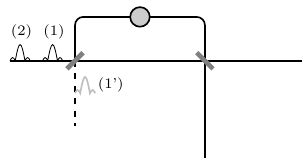}
\end{minipage}
&
\begin{minipage}[r]{\linewidth}
A general single-photon qubit,
		$\alpha\fet{0{,}1}+\beta\fet{1{,}0}$,
		enters the interferometer (modes (2) and (1)).
		Bob adds a vacuum ancilla (1') that 
		interferes with mode (1) at the first beam splitter ($\BS_1$).
\end{minipage}
\end{tabular}
\begin{tabular}{p{0.4\columnwidth} p{0.55\columnwidth}}
\begin{minipage}[c]{\linewidth}
\centering
\includegraphics{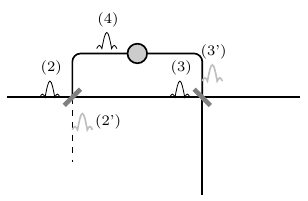}
\end{minipage}
&
\begin{minipage}[r]{\linewidth}
Pulses (1) and (1') 
			interfere and yield pulses (3) and (4)
			 in the short arm and the long arm, respectively:
			$\alpha\fet{0}_{1'}\fet{1}_1 \mapm{\BS_1} 
			\frac{\alpha}{\sqrt{2}} 
			(\fet{0}_{4}\fet{1}_3 + i\fet{1}_{4}\fet{0}_3)$.
			Pulse (3) is about to enter $\BS_2$, 
			so a vacuum ancilla (3') is added. 
			Pulse (2) is about to enter $\BS_1$, so
			a vacuum ancilla (2') is added. 
\end{minipage}
\end{tabular}
\begin{tabular}{p{0.4\columnwidth} p{0.55\columnwidth}}
\begin{minipage}[c]{\linewidth}
\centering
\includegraphics{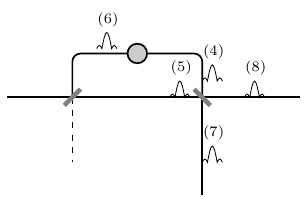}
\end{minipage}
&
\begin{minipage}[r]{\linewidth}
Pulses (7) and (8) 
            are created by the interference of
            (3) and (3'): $\frac{\alpha}{\sqrt{2}}\fet{1}_3\fet{0}_{3'}
 \mapm{\BS_2}\frac{\alpha}{2} (i\fet{0}_{8}\fet{1}_7
			+\fet{1}_{8}\fet{0}_7)$.
			Pulses (5) and (6) are created by the 
			interference of (2) and (2') in $\BS_1$:
			$\beta\fet{0}_{2'}\fet{1}_2 \mapm{\BS_1}
			\frac{\beta}{\sqrt{2}} (\fet{0}_{6}\fet{1}_5 + 
			i\fet{1}_{6}\fet{0}_5)$.
\end{minipage}
\end{tabular}
\begin{tabular}{p{0.4\columnwidth} p{0.55\columnwidth}}
\begin{minipage}[c]{\linewidth}
\centering
\includegraphics{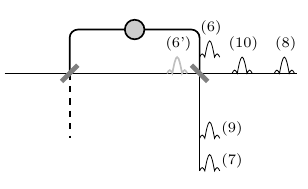}
\end{minipage}
&
\begin{minipage}[r]{\linewidth}
Pulses (9) and (10) are 
		created by the interference of (4)
		and (5) in the second beam splitter: 
		$ \frac{i\alpha}{\sqrt{2}}\fet{0}_5\fet{1}_4
		+ \frac{\beta}{\sqrt{2}}\fet{1}_5\fet{0}_4 
		\mapm{\BS_2} 
		  \frac{i(\alpha+\beta)}{2}\fet{0}_{10}\fet{1}_9
		+ \frac{\beta-\alpha}{2} \fet{1}_{10}\fet{0}_9$ . 
		Pulse (6) is about to enter $\BS_2$,
		so a vacuum ancilla is added (6').
\end{minipage}
\end{tabular}
\begin{tabular}{p{0.4\columnwidth} p{0.55\columnwidth}}
\begin{minipage}[c]{\linewidth}
\centering
\includegraphics{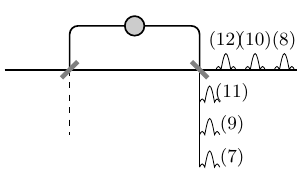}
\end{minipage}
&
\begin{minipage}[r]{\linewidth}
Pulses (11) and (12) are 
			created by the interference of (6)
			and (6') in $\BS_2$:
			$\frac{i\beta}{\sqrt{2}}\fet{0}_{6'}\fet{1}_{6}
\mapm{\BS_2}
			\frac{\beta}{2} (i \fet{0}_{12}\fet{1}_{11} - 
			\fet{1}_{12}\fet{0}_{11})$.
\end{minipage}
\end{tabular}
\end{framed}
\caption{\small
		Evolution in time of two modes 
		through the interferometer with $\phi=0$:  
		$\fet{0{,}0{,}0{,}0}_{6',3',2',1'}
        \left (\alpha\fet{0}_2\fet{1}_1 + 
        \beta\fet{1}_2\fet{0}_1 \right)
        \to 
        \bigl (\frac{\alpha}{2}\fet{0{,}0{,}0{,}0{,}0{,}1}+\frac{\beta-\alpha}{2} \fet{0{,}0{,}0{,}0{,}1{,}0}
        -\frac{\beta}{2}\fet{0{,}0{,}0{,}1{,}0{,}0} 
        + \frac{i\alpha}{2}\fet{0{,}0{,}1{,}0{,}0{,}0}
        + \frac{i(\alpha+\beta)}{2}\fet{0{,}1{,}0{,}0{,}0{,}0}
        + \frac{i\beta}{2} \fet{1{,}0{,}0{,}0{,}0{,}0} \bigr)_{11, 9, 7, 12, 10, 8}$.
		The output state is denoted by modes
		 $\fet{n_{d_2}, n_{d_1}, n_{d_0}, n_{s_2}, n_{s_1}, n_{s_0}}$.
}
\Description{fully described inside the figure.}
\label{fig:2evo}
\end{figure}

We are now ready to consider the setup of 
Figure~\ref{fig:lab-xy} and two input modes, $t'_0$ and $t'_1$,
that enter the interferometer one after the other, 
with exactly the same time difference $\Delta T$ as the interferometer's arms.
As a result of this precise timing, the two modes are transformed into 
a superposition of only six modes
(instead of eight modes) at the outputs (see Figure~\ref{fig:2evo}).
Four (vacuum state) ancillas are added during the process 
and the resulting six modes are $t_0$, $t_1$, $t_2$ at the `s' arm 
and the `d' arm of the interferometer. 
A basis state of this Fock-space is, therefore,
$\fet{n_{d_2}, n_{d_1}, n_{d_0}, n_{s_2}, n_{s_1}, n_{s_0}}$.
If exactly one photon enters the interferometer,
we can  
use Eq.~\eqref{eqn:single-pulse-evolution}
to obtain 
\begin{align}\nonumber
\fet{0{,}0{,}0{,}0}
\fet{0}_{t'_1}
\fet{1}_{t'_0} &\mapsto 
(\fet{0{,}0{,}0{,}0{,}0{,}1}-e^{i\phi}\fet{0{,}0{,}0{,}0{,}1{,}0}+i\fet{0{,}0{,}1{,}0{,}0{,}0} +i e^{i\phi}\fet{0{,}1{,}0{,}0{,}0{,}0}) \thickspace / 2\text{ ,} \\
\fet{0{,}0{,}0{,}0}
\fet{1}_{t'_1}
\fet{0}_{t'_0} &\mapsto 
(\fet{0{,}0{,}0{,}0{,}1{,}0}-e^{i\phi}\fet{0{,}0{,}0{,}1{,}0{,}0}+i\fet{0{,}1{,}0{,}0{,}0{,}0} +i e^{i\phi}\fet{1{,}0{,}0{,}0{,}0{,}0}) \thickspace / 2\text{ .}
\end{align}
Recall that  $\ket{0}=\fet{0{,}1}_{t'_1,t'_0}$ and $\ket{1}=\fet{1{,}0}_{t'_1,t'_0}$.
It follows that an arbitrary qubit is transformed as
\begin{equation}
    \begin{array}{lllll}
        \fet{0{,}0{,}0{,}0} \left( \alpha\fet{0{,}1} + \beta\fet{1{,}0} \right) & \longrightarrow \bigg( &\frac{\alpha}{2}\fet{0{,}0{,}0{,}0{,}0{,}1} & + & \frac{\beta-\alpha e^{i\phi}}{2} \fet{0{,}0{,}0{,}0{,}1{,}0} \\
        && -  \frac{\beta e^{i\phi}}{2}\fet{0{,}0{,}0{,}1{,}0{,}0} & + & \frac{i\alpha}{2}\fet{0{,}0{,}1{,}0{,}0{,}0} \\
        && +  \frac{i(\alpha e^{i\phi}+\beta)}{2}\fet{0{,}1{,}0{,}0{,}0{,}0} & + & \frac{i\beta e^{i\phi}}{2} \fet{1{,}0{,}0{,}0{,}0{,}0} \bigg)\text{ .}
    \end{array}
\end{equation}

\subsection*{Reversal of Single Photon Through Mach-Zehnder Interferometer}
In our work, we sometimes wish to see the time-reversed operation of a device on
a certain measured state, in order to see which states affect it and in which way.
We will now perform this analysis for the Mach-Zehnder interferometer.

While the inversion of a unitary transformation is not complicated, it requires two things:
a characterization of the relevant Hilbert space, and a characterization of its action on
all basis states of that space.

Let $U$ denote the unitary action that the interferometer performs.
Let us denote the non-blocked and blocked input arms (a) and (b) respectively,
and let us denote the ``straight'' and ``down'' output arms (s) and (d) respectively.

While a pulse can arrive at the interferometer at any instance of time (which is continuous),
we know that pulses only interfere with each other if their time difference is equal to the time difference
of the interferometer's two paths.
Thus, we define some instance of time as $t=0$, and consider all times $\{t_n\}_n = \{n\cdot\Delta T\ \;|\; n \in \mathbb{Z}\}$, where $\Delta T$ is the
interferometer's time difference.

The input space is thus:
\begin{equation}
	\spn\{\ket{a_n}, \ket{b_n} \;|\; n \in \mathbb{Z}\},
\end{equation}
and the output space is
\begin{equation}
	\spn\{\ket{s_n}, \ket{d_n} \;|\; n \in \mathbb{Z}\},
\end{equation}
while we have already shown that the evolution for $\ket{a_n}$ is given by:
\begin{equation}
	U\ket{a_n} = \left(\ket{s_n} + i\ket{d_n} - e^{i\phi}\ket{s_{n+1}} + i e^{i\phi}\ket{d_{n+1}}\right) / \thickspace 2.
\end{equation}

We now need to analyze the evolution of a pulse from the \textit{blocked} arm,
as a time-reversed pulse from one of the output arms could quite possibly reach it.

\begin{equation}
	\begin{array}{rcl}
		\ket{b_n} &\underset{\text{beam splitter}}{\mapsto}& \left(i\ket{\mathrm{short}_n} + \ket{\mathrm{long}_n}\right) / \thickspace \sqrt{2}\\
		&\underset{\text{phase + delay}}{\mapsto}& \left(i\ket{\mathrm{short}_n} + e^{i\phi}\ket{\mathrm{long}_{n+1}}\right) / \thickspace \sqrt{2}\\
		&\underset{\text{beam splitter}}{\mapsto}& \left(i\ket{s_n} - \ket{d_n} + i e^{i\phi}\ket{s_{n+1}} + e^{i\phi}\ket{d_{n+1}}\right) / \thickspace 2. \\
	\end{array}
\end{equation}

Thus, $U$ can be written in matrix form from $(\ket{a_{-1}}, \ket{b_{-1}}, \ket{a_0}, \ket{b_0}, \ket{a_1}, \ket{b_1}, \ldots)$ \\
to $(\ket{s_{-1}}, \ket{d_{-1}}, \ket{s_0}, \ket{d_0}, \ket{s_1}, \ket{d_1}, \ldots)$ as:

\begin{equation}
	U = \frac{1}{2}
	\begin{pmatrix}
		1 & i & 0 & 0 & \cdots \\
		i & -1 & 0 & 0 \\
		-e^{i\phi} & i e^{i\phi} & 1 & i \\
		i e^{i\phi} & e^{i\phi} & i & -1 \\
		0 & 0 & -e^{i\phi} & i e^{i\phi} \\
		0 & 0 & i e^{i\phi} & e^{i\phi} \\
		\vdots & & &
	\end{pmatrix}.
\end{equation}

By inverting the matrix, we get:

\begin{equation}
	U^{-1} = U^\dagger = \frac{1}{2}
	\begin{pmatrix}
		1 & -i & -e^{-i\phi} & -i e^{-i\phi} & 0 & 0 & \cdots \\
		-i & -1 & -i e^{-i\phi} & e^{-i\phi} & 0 & 0 \\
		0 & 0 & 1 & -i & -e^{-i\phi} & -i e^{-i\phi} \\
		0 & 0 & -i & -1 & -i e^{-i\phi} & e^{-i\phi} \\
		\vdots
	\end{pmatrix}.
\end{equation}

Finally, the formula form for each basis element is given by:
\begin{equation}
	\begin{array}{lcl}
		U^\dagger \ket{s_n} &=& -e^{-i\phi}\ket{a_{n-1}} - i e^{-i\phi}\ket{b_{n-1}} + \ket{a_n} - i\ket{b_n}, \\
		U^\dagger \ket{d_n} &=& -i e^{-i\phi}\ket{a_{n-1}}  + e^{-i\phi}\ket{b_{n-1}} - i\ket{a_n} - \ket{b_n}. \\
	\end{array}
    \label{eqn:backward-pulse}
\end{equation}

\section{Example: Insecurity of Interferometric BB84}
\label{app:rsa_main_example}
In this appendix, we show how to employ the Reversed-Space method 
on an interferometric BB84 scheme and obtain the attack mentioned in Section~\ref{subsec:RSA-example}.
 
In interferometric BB84 schemes, a qubit is encoded using a single photon in two possible time-bins.
One basis can be defined by the photon arriving at one time or the other.
Other qubit states are superpositions of these two times.
When only states that use both times in a superposition are used, the scheme is called \textit{time-multiplexed, phase-encoded}.
Such schemes were implemented by Townsend~\cite{Townsend1994}
and many others (e.g.,~\cite{Hughes2000,Kimura2004,Gobby2004,Yuan2009,Lucamarini2013,Yuan2018}; see also~\cite{Gisin2002, Xu2020}).

In order to produce and measure pulses with time superposition, it is common to use 
\textit{interferometers}
(see below and Appendix~\ref{app:quantum_optics}).
Yet, once a protocol is implemented using photons and interferometers, there are two immediate reasons for an expansion of the quantum space in use. First, interferometers 
inherently introduce a higher-dimensional space. 
Second, having pulses with zero or more than one photon also implies a higher dimension.\footnote{There are also other possible causes for space enlargement. For example, this could happen due to imperfect generation or measurement of the pulse shape in either the frequency or time domain.}

In this appendix, we demonstrate a Reversed-Space Attack on an interferometric BB84 implementation (described below).
The attack we define here exposes a vulnerability inherent to a large class of implementations,
and is \textit{directly applicable} to the implementations in~\cite{Walton2003,Nambu2004,Jaeger2006} and the NICT-NEC implementation in~\cite{Sasaki2011}.

Our focus in this appendix is on a specific physical apparatus.
Note that not all physical implementations of interferometric QKD are insecure in this fashion~\cite{gelles_security_2012}, 
but other implementation methods should still be analyzed~\cite{Boaron2018,Beutel2021}.

We begin by describing the protocol implementation and the setup Bob uses.

\enlargethispage{-2ex}

\subsection{Interferometric Implementation of BB84}
\label{subsec:InterferoBB84}
Consider an implementation of BB84 that uses two time-separated modes (pulses) to encode a qubit.
For every transmission, the first mode arrives to Bob's lab at time $t_0'$, and the second
mode at $t_1'=t_0'+\Delta T$. We denote these pulses as $\ket{t'_0}$ and $\ket{t'_1}$, respectively.
The ideal Alice sends one of the following four states,
\begin{align*}
\ket{0}_{\sA} \equiv 
\ket{t'_0}  
&&
\ket{+}_{\sA} \equiv 
\left ( \ket{t'_0} + \ket{t'_1} \right )/\sqrt{2}  \phantom{\text{ ,}}
\\
\ket{1}_{\sA} \equiv 
\ket{t'_1}
&&
\ket{-}_{\sA} \equiv 
\left ( \ket{t'_0} - \ket{t'_1} \right )/\sqrt{2}  \text{ ,} 
\end{align*}
where $\{\ket{0}_{\sA},\ket{1}_{\sA}\}$ is the computational basis, and $\{\ket{+}_{\sA},\ket{-}_{\sA}\}$ is the Hadamard basis.

Bob measures the qubit using a Mach-Zehnder interferometer, which is a device composed of
two beam splitters (BS) with one short path, one long path, and a 
controlled phase shifter
$P_\phi$, that is placed at the long arm of the interferometer. 
(See Appendix~\ref{app:quantum_optics} for details on linear-optics devices
and an analysis of their operation on photonic states).
The length difference between the two arms is determined by $\Delta T$: 
when the first pulse travels through the long arm, and the second through
the short arm, they arrive together at the output.
Due to that exact timing of the pulses,   
each incoming qubit  
is transformed into a superposition of \textit{six} possible modes:
three time modes ($t_0$, $t_1$, $t_2$) at the straight ($s$) output arm
of the interferometer, and
three modes at the down~($d$) output arm; see Figure~\ref{fig:lab-xy}.

For simplicity, we denote these modes as $s_0, s_1, s_2, d_0, d_1, d_2$.
In order to construct this attack, it is sufficient to only consider photons with zero or one photon.
Thus, we use the states
we can use the states $\{\ket{s_j}$, $\ket{d_j} : j \in \{0,1,2\}\}$
along with the vacuum state $\ket{V}$ which denotes a pulse that has no photons in any of the modes.\footnote{Using the Fock-space notations (Section~\ref{subsec:fock_notation})
and the description of interferometers in Appendix~\ref{app:quantum_optics}, 
a basis state in Bob's space is  $\fet{n_{d_2}, n_{d_1}, n_{d_0}, n_{s_2}, n_{s_1}, n_{s_0}}$, and we define
$\fet{0{,}0{,}0{,}0{,}0{,}1} \equiv \ket{s_0}$; 
$\fet{0{,}0{,}0{,}0{,}1{,}0} \equiv \ket{s_1}$; 
$\fet{0{,}0{,}0{,}1{,}0{,}0} \equiv \ket{s_2}$; 
$\fet{0{,}0{,}1{,}0{,}0{,}0} \equiv \ket{d_0}$; 
$\fet{0{,}1{,}0{,}0{,}0{,}0} \equiv \ket{d_1}$;
$\fet{1{,}0{,}0{,}0{,}0{,}0} \equiv \ket{d_2}$, and the 
vacuum state $\fet{0{,}0{,}0{,}0{,}0{,}0} \equiv \ket{V}$.}

As shown in Appendix~\ref{app:quantum_optics} interferometer evolves these states
according to
$\ket{V}\mapsto\ket{V}_{\sB}$ and
\begin{equation}\label{eqnB+-On01}
\begin{split}
\ket{t'_0}   &\mapsto 
 (\ket{s_0}_{\sB}-e^{i\phi}\ket{s_1}_{\sB}+i\ket{d_0}_{\sB}
	+i e^{i\phi}\ket{d_1}_{\sB})  / 2 \\  
\ket{t'_1}   &\mapsto 
 (\ket{s_1}_{\sB}-e^{i\phi}\ket{s_2}_{\sB}+i\ket{d_1}_{\sB}
	+i e^{i\phi}\ket{d_2}_{\sB}) / 2.
\end{split}
\end{equation}

In our example, Bob fixes the phase $\phi$ to 0, regardless of the basis in which he wishes to measure; thus, $U_{\sB}$ is fixed and identical to the operation of the interferometer.
Thus, Alice's qubit evolves in the interferometer as  
\begin{equation}
\begin{array}{rcl}
\label{eqnB+-On+-}
\ket{0}_{\sA}   &\mapsto &
 (\ket{s_0}_{\sB}-\phantom{2}\ket{s_1}_{\sB}  \phantom{{}-\ket{s_1}_{\sB}} +i\ket{d_0}_{\sB}  	+\phantom{2}i\ket{d_1}_{\sB} \phantom{{}+i\ket{d_2}_{\sB}}
 )  / 2 \\  
\ket{1}_{\sA}    &\mapsto &
 ( \phantom{\ket{s_0}_{\sB}+2}\ket{s_1}_{\sB}-\ket{s_2}_{\sB}  \phantom{{}+i\ket{s_1}_{\sB}}     +\phantom{2} i\ket{d_1}_{\sB}
	+i\ket{d_2}_{\sB}) / 2 \\ [0.5em]
\ket{+}_{\sA}  &\mapsto  &
 (\ket{s_0}_{\sB}  \phantom{{}-2\ket{s_1}_{\sB}} -\ket{s_2}_{\sB} 
 +i\ket{d_0}_{\sB} +2i \ket{d_1}_{\sB} +i\ket{d_2}_{\sB}) \thickspace / \sqrt{8}\\
\ket{-}_{\sA}  &\mapsto &
 (\ket{s_0}_{\sB} -2 \ket{s_1}_{\sB} +\ket{s_2}_{\sB} 
+i\ket{d_0}_{\sB} \phantom{{}+2i\ket{d_1}_{\sB}} -i\ket{d_2}_{\sB}) \thickspace / \sqrt{8}  .
\end{array}
\end{equation}

In order to measure the Hadamard basis, 
Bob opens his detectors at time $t_1$ at both arms.
A click at the ``down'' direction 
(i.e., measuring the state $\ket{d_1}$) 
means the bit-value $0$, while
a click at the ``straight'' direction ($\ket{s_1}$) means $1$.
The other modes are considered as a
loss (namely, they are not measured)
since they do not reveal the value of the original qubit.

Similarly, in order to measure in the computational basis, Bob need not measure
time $t_1$ as it does not reveal the value of the original bit. 
Bob may open his detector in times $t_0, t_2$ (on both hands) where the former implies measurement of the bit~$0$ and
the latter implies measurement of the bit $1$.

\subsection{Identifying the Reversed Space of the Interferometric Setup}
\label{subsec:spaces-example}

We now follow the framework defined in Section~\ref{sec:RSA} and specify 
the space Bob measures and its reversed space.
We then derive the corresponding reversed space that applies in each setting.

Let us first analyze what states affect a single mode measured by Bob.
As derived in Appendix~\ref{app:quantum_optics} (Eq.~\eqref{eqn:backward-pulse}), a reversal of a single mode through the interferometer (i.e., $U_{\sB}^\dagger$) is given by
\begin{align}\label{eqn:Udag}
\begin{split}
\ket{s_n}   &\mapsto
 (\phantom{-i}\ket{a_n}-\phantom{i}e^{-i\phi}\ket{a_{n-1}}-i\ket{b_n} -i e^{-i\phi}\ket{b_{n-1}})  / 2  \\
\ket{d_n}  &\mapsto 
 (-i\ket{a_n} - i e^{-i\phi}\ket{a_{n-1}}-\phantom{i}\ket{b_n} + \phantom{i}e^{-i\phi}\ket{b_{n-1}})/2\text{,}
 \end{split}
 \end{align}
where $\ket{a_n}$ is a pulse in the input arm of the interferometer at time~$t_n$, and $\ket{b_n}$ is a pulse in the blocked arm of the interferometer
at time~$t_n$.

Now consider the six modes that Bob potentially measures in every BB84 transmission, i.e.,  time-bins $t_0,t_1,t_2$ in both output arms. 
Bob's measured space $\mathcal{H}^{\sB}$ is the span of 
$\{\ket{V}$, $\ket{d_0}$, $\ket{d_1}$, $\ket{d_2}$, $\ket{s_0}$, $\ket{s_1}$, $\ket{s_2} \}$. 
We now use the reversed transformation in Eq.~\eqref{eqn:Udag}
to derive the reversed space~$\mathcal{H}^{\sP}$.
After ``tracing out'' the blocked ancillary system (the ``$b$'' arm), we get that 
$\mathcal{H}^{\sP}$ is the space
that allows the photon to be in any superposition of time modes $t'_{-1}$ to~$t'_2$ of the interferometer input (the ``$a$'' arm). 
Surprisingly, this space is much larger than~$\mathcal{H}^{\sA}$.

Now, our  analysis can only focus on the space spanned by
$\{ \ket{V}$, $\ket{t'_{-1}}$, $\ket{t'_{0}}$, $\ket{t'_{1}}$, $\ket{t'_{2}}\}$.

\subsection{A Reversed-Space Attack}
\label{subsec:attack6}
We now design an oblivious 
attack on the BB84 realization described in Section~\ref{subsec:InterferoBB84}, using the reversed-space methodology described in Section~\ref{subsec:RSA_oblivious}.

As mentioned above,
Bob's unitary $\calu_{\sB}$ is the same for both the computational and the Hadamard bases
(that is, $\beta^{\mathrm{c}}=\beta^{\mathrm{H}}$) and is characterized by 
\[
\beta^{\mathrm{H}}_{\begin{array}[t]{l} \scriptstyle k=\{t'_{-1},t'_0,t'_1,t'_2\}, \\[-3pt]
\scriptstyle j=\{ s_0, s_1, s_2, d_0, d_1,d_2\}\end{array}} \!\!\!\!\!\!=
\frac{1}{2}\left (
    \begin{array}{cccccc}
 -1  &  0   &  0  & i   &  0  &  0 \\
 1   &  -1  &  0  & i   &  i  &  0 \\
 0   &  1  &  -1  &  0  &  i  &  i \\
 0   &  0  &   1  &  0  &  0  &  i
    \end{array}
\right ),
\]
which is immediately obtained by extending Eq.~\eqref{eqnB+-On01} with $\phi=0$ to times $t'_{-1}$ and~$t'_2$.

Denote with $B = \{\ket{V}$, $\ket{d_0}$,
$\ket{d_1}$, $\ket{d_2}$, $\ket{s_0}$, $\ket{s_1}$, $\ket{s_2} \}$
a basis of Bob's measured space~$\mathcal{H}^{\sB}$. 
When Bob measures in the Hadamard basis, he interprets his measurement in the following way,
$J_0 = \{ \ket{d_1}\}$;
$J_1 = \{ \ket{s_1}\}$;
$J_\loss = B \setminus (J_0 \cup J_1)$; and 
$J_\inv = \emptyset$.\footnote{We limit the analysis to pulses that include at most a single photon. Under this assumption, there are no invalid states for this setting.}
Consider the case where Alice sends $\ket{+}$,
namely, $\alpha_{t'_0}=\alpha_{t'_1}=\frac{1}{\sqrt{2}}$.
An error occurs if Bob measures $\ket{s_1}$, $J_\err =   \{ \ket{s_1} \} $,
and by Observation~\ref{clm:ZeroErrorRequirement}, our attack is required to satisfy
\begin{equation}\label{eqn:ReqRobX1}
 -\frac{1}{2\sqrt{2}}(\epsilon_{t'_0,t'_0}\ket{E_{t'_0,t'_0}}_{\sE}+
   \epsilon_{t'_1,t'_0}\ket{E_{t'_1,t'_0}}_{\sE}) 
    +\frac{1}{2\sqrt{2}}(\epsilon_{t'_0,t'_1}\ket{E_{t'_0,t'_1}}_{\sE}+
   \epsilon_{t'_1,t'_1}\ket{E_{t'_1,t'_1}}_{\sE}) = 0.
\end{equation} 
Similarly, when Alice sends $\ket{-}$ an error happens when Bob measures $J_\err=\{\ket{d_1}\}$, and thus we require that
\begin{equation} 
\label{eqn:ReqRobX2}
\frac{i}{2\sqrt{2}}(\epsilon_{t'_0,t'_0}\ket{E_{t'_0,t'_0}}_{\sE}-\epsilon_{t'_1,t'_0}
\ket{E_{t'_1,t'_0}}_{\sE})
+\frac{i}{2\sqrt{2}}(\epsilon_{t'_0,t'_1}\ket{E_{t'_0,t'_1}}_{\sE}-\epsilon_{t'_1,t'_1}
\ket{E_{t'_1,t'_1}}_{\sE})
 = 0 \text{.}
\end{equation}

As for the computational basis, 
Bob interprets his outcome according to 
$J_0 = \{ \ket{d_0}, \ket{s_0}\}$,
$J_1 = \{ \ket{d_2}, \ket{s_2}\}$,
$J_\inv = \emptyset$, and
$J_\loss  = B \setminus ( J_0 \cup J_1)$. 
Following Observation~\ref{clm:ZeroErrorRequirement},
an attack $\calu_{\sE}$ which causes no errors is required to satisfy
\begin{equation}
\begin{split}
i\epsilon_{t'_0,t'_1}\ket{E_{t'_0,t'_1}} +i\epsilon_{t'_0,t'_2}\ket{E_{t'_0,t'_2}}=0, \quad\quad\quad
-\epsilon_{t'_0,t'_1}\ket{E_{t'_0,t'_1}} +\epsilon_{t'_0,t'_2}\ket{E_{t'_0,t'_2}} =0, 
\end{split}
\end{equation}
corresponding to the case where Alice sends $\ket{0}$, i.e.\ $\alpha_{t'_0} = 1$, $\alpha_{t'_1} =0$,
and $J_\err = \{ \ket{d_2}, \ket{s_2} \}$, as well as
\begin{equation}
\begin{split}
i\epsilon_{t'_1,t'_{-1}}\ket{E_{t'_1,t'_{-1}}} + i\epsilon_{t'_1,t'_0}\ket{E_{t'_1,t'_0}} =0, \quad\quad\quad
-\epsilon_{t'_1,t'_{-1}}\ket{E_{t'_1,t'_{-1}}}+ \epsilon_{t'_1,t'_0}\ket{E_{t'_1,t'_0}}=0,  
\end{split}
\end{equation}
corresponding the case where Alice sends  $\ket{1}$, 
i.e.\ $\alpha_{t'_0} = 0$, $\alpha_{t'_1} =1$, and $J_\err = \{ \ket{d_0}, \ket{s_0} \}$.
This leads to the constraints
$\epsilon_{t'_0,t'_1}=\epsilon_{t'_0,t'_2}=0$
and $\epsilon_{t'_1,t'_{-1}}=\epsilon_{t'_1,t'_0}=0$.

Combining all the above requirements yields that 
the only possible attacks are of the form
\begin{equation}\label{eqn:attackXZ}
\begin{split}
\ket{0}_{\sE}\ket{0}_{\sA} &\mapm{\calu_{\sE}} 
  p \ket{\phi}_{\sE}\ket{t'_0}_{\sP}+ p_1\ket{\phi_1}_{\sE}\ket{t'_{-1}}_{\sP}+ p_2\ket{\psi_0}_{\sE}\ket{V}_{\sP}, 
\\ %
\ket{0}_{\sE}\ket{1}_{\sA} &\mapm{\calu_{\sE}} 
  p \ket{\phi}_{\sE}\ket{t'_1}_{\sP} +p_3\ket{\phi_2}_{\sE}\ket{t'_2}_{\sP}+ p_4\ket{\psi_1}_{\sE}\ket{V}_{\sP},
\end{split}
\end{equation}
with $|p|^2+ |p_1|^2+|p_2|^2 = |p|^2+ |p_3|^2+|p_4|^2 =1$.
Using Eq.~\eqref{eqn:attackXZ} it is easy to devise an attack and demonstrate that the protocol
is completely insecure in the sense that there exists an attack that leaks information without causing any errors.
For instance, let 
\begin{equation}\nonumber
\ket{0}_{\sE}\ket{0}_{\sA} \mapm{\calu_{\sE}} \ket{E_1}_{\sE}\ket{t'_{-1}}_{\sP},  \qquad \qquad
 \ket{0}_{\sE}\ket{1}_{\sA} \mapm{\calu_{\sE}} \ket{E_2}_{\sE}\ket{t'_2}_{\sP}, 
\end{equation}
with orthogonal $\ket{E_1}$, $\ket{E_2}$. 
As shown in Section~\ref{sec:applicability}, this attack is related to the ``faked states'' attack family~\cite{makarov_faked_2005,makarov_faked_2008} described in Section~\ref{subsubsec:qkd_exploits}.

While the above attack never
causes an error, it increases the loss rate---Bob always gets a loss when using the Hadamard basis.
This means that only bits encoded using the computational basis are used for transferring information,
and Eve can copy the information, thus the scheme is insecure.
We can compose another attack that does not have the property of causing a loss-rate~$1$ in a specific basis.
For instance, by letting~$p>0$ Eve does not force a loss in the Hadamard basis,
yet she does not learn the information for that basis.

\section{Interferometric BB84 With Added Measurements}
\label{app:rsa_app_added_measurements}
In this appendix, we revisit the attack on the interferometric BB84 implementation from Appendix~\ref{app:rsa_main_example}, in which Eve uses time modes other than $t'_0$ and $t'_1$ in order to fool Bob's detection device. 

However, by doing so, Eve creates pulses at times $t_3$ and $t_{-1}$ at Bob's device, while such modes can never happen if only the incoming qubits arrive in time mode $t'_0$ and $t'_1$, as they should.

Therefore, let us consider a defense mechanism against the Reversed-Space Attack presented in Appendix~\ref{app:rsa_main_example}.
Bob will measure both arms at times $t_{-1}$ and $t_3$ in addition to the other six modes he measured before.
Any pulse coming in any of these new modes will be interpreted by Bob as an invalid result.

We can now use the Reversed-Space formalism to analyze this new implementation, define the enlarged space that Eve can use (note, this space will be different form the one in Appendix~\ref{app:rsa_main_example} due to the added measurements), and check if Reversed-Space Attacks still exist in this new setting.

We constrain ourselves in the following ways:
(i) Eve can only send zero or one photons; (ii) Eve attacks each pulse separately; and (iii) the optical components of Bob's device are perfectly implemented.

\subsection{Bob's Measured Space}
As stated, Bob will now measure pulses arriving at times $t_{-1}$ and $t_3$. Thus, his measured space is:
\begin{equation}
	\mathcal{H}^{\sB} = \spn \left\{ \ket{V}, \ket{s_{-1}}, \ket{d_{-1}}, \ket{d_0}, \ket{d_0}, \ket{s_1}, \ket{d_1}, \ket{s_2}, \ket{d_2}, \ket{s_3}, \ket{d_3} \right\}.
\end{equation}

Bob's unitary is the same for both measurement bases, and is identical to the one given in Eq.~\eqref{eqn:mzi_generic_single_photon} ($\theta = 0$ still holds):
\begin{equation}
	\calu_{\sB} \ket{t'_n} = \left( \ket{s_n} + i\ket{d_n} - \ket{s_{n+1}} + i\ket{d_{n+1}} \right) / \thickspace 2.
\end{equation}

Bob's interpretation sets for the computational basis are given by:
\begin{equation}
	\begin{array}{rcl}
		J_0 &=& \left\{ \ket{s_0}, \ket{d_0} \right\}, \\
		J_1 &=& \left\{ \ket{s_2}, \ket{d_2} \right\}, \\
		J_{\loss} &=& \left\{ \ket{d_1}, \ket{s_1} \right\}, \\
		J_{\inv} &=& \left\{ \ket{s_{-1}}, \ket{d_{-1}}, \ket{s_3}, \ket{d_3} \right\} 		.
	\end{array}
\end{equation}

Bob's interpretation sets for the Hadamard basis are given by:
\begin{equation}
	\begin{array}{rcl}
		J_0 &=& \left\{ \ket{d_1} \right\}, \\
		J_1 &=& \left\{ \ket{s_1} \right\}, \\
		J_{\loss} &=& \left\{ \ket{s_0}, \ket{d_0}, \ket{s_2}, \ket{d_2} \right\}, \\
		J_{\inv} &=& \left\{ \ket{s_{-1}}, \ket{d_{-1}}, \ket{s_3}, \ket{d_3} \right\} 	.	
	\end{array}
\end{equation}

\subsection{Eve's Attack}

As shown in Appendix~\ref{app:rsa_main_example}, each of Bob's measured space is affected by its respective
time bin at the entrance to his apparatus, as well as the one preceding it.
From this argument, the reversed space is given by:
\begin{equation}
	\mathcal{H}^{\sP} = \spn \{ \ket{t'_{-2}}, \ket{t'_{-1}} \ket{t'_0}, \ket{t'_1}, \ket{t'_2}, \ket{t'_3}\}.
\end{equation}

Thus, we can define Eve's general form of attack (without the required obliviousness constraints) as:
\begin{equation}
	\begin{array}{rcccl}
		\calu_{\sE} \ket{0}_{\sE} \ket{\fzero}_{\sA} &=& \epsilon_{0,-2} \ket{E_{0,-2}} \ket{t'_{-2}} &+& \epsilon_{0,-1} \ket{E_{0,-1}} \ket{t'_{-1}} \\
		&+& \epsilon_{0,0} \ket{E_{0,0}} \ket{t'_0} &+& \epsilon_{0,1} \ket{E_{0,1}} \ket{t'_1} \\
		&+& \epsilon_{0,2} \ket{E_{0,2}} \ket{t'_2} &+& \epsilon_{0,3} \ket{E_{0,3}} \ket{t'_3}, \\
		\calu_{\sE} \ket{0}_{\sE} \ket{\fone}_{\sA} &=& \epsilon_{1,-2} \ket{E_{1,-2}} \ket{t'_{-2}} &+& \epsilon_{1,-1} \ket{E_{1,-1}} \ket{t'_{-1}} \\
		&+& \epsilon_{1,0} \ket{E_{1,0}} \ket{t'_0} &+& \epsilon_{1,1} \ket{E_{1,1}} \ket{t'_1} \\
		&+& \epsilon_{1,2} \ket{E_{1,2}} \ket{t'_2} &+& \epsilon_{1,3} \ket{E_{1,3}} \ket{t'_3}. \\
	\end{array}
\end{equation}

\subsection{Attack Constraints}
The constraints that would yield an oblivious attack according to Observation~\ref{clm:ZeroErrorRequirement}, 
are given by:
$$ \forall \ket{\psi}_{\sA} = \sum_i \alpha_i \ket{i}_{\sA},\quad
\forall \calu_{\sB_s} = \sum_{k,j} \beta_{k,j}^s \ket{j}_{\sB}\bra{k}_{\sA},\quad
\forall \ket{j} \in J_{\err} :
$$ $$ \sum_{i,k}\alpha_{i} \epsilon_{i,k}\beta^s_{k,j} \ket{E_{i,k}}_{\sE}=0. $$

Together with normalization and orthogonality conditions.
Let us apply the constraints to each measurement basis and basis vector separately.

\paragraph{Computational Basis}
Consider Alice's state $\ket{\psi}_{\sA} = \ket{\fzero} = \ket{t'_0}$.
The erroneous states that must be avoided are as follows:
\begin{equation}
	J_{\err} = J_1 \cup J_{\inv} = \left\{ \ket{s_{-1}}, \ket{s_2}, \ket{s_3}, \ket{d_{-1}}, \ket{d_2}, \ket{d_3} \right\}.
\end{equation}

\begin{itemize}
	\item For $\ket{j} = \ket{s_{-1}}$:
	\begin{equation}
	\label{eqn:defense_z_0_-2}
		\epsilon_{0,-2}\ket{E_{0,-2}} - \epsilon_{0,-1} \ket{E_{0,-1}} = 0.
	\end{equation}

	\item For $\ket{j} = \ket{s_2}$:
	\begin{equation}
		\epsilon_{0,1} \ket{E_{0,1}} - \epsilon_{0,2} \ket{E_{0,2}} = 0.
	\end{equation}

	\item For $\ket{j} = \ket{s_3}$:
	\begin{equation}
		\begin{array}{c}
			\epsilon_{0,2} \ket{E_{0,2}} - \epsilon_{0,3} \ket{E_{0,3}} = 0 \\
			\Downarrow \\
			\epsilon_{0,1} \ket{E_{0,1}} = \epsilon_{0,2} \ket{E_{0,2}} = \epsilon_{0,3} \ket{E_{0,3}}.
		\end{array}
	\end{equation}

	\item For $\ket{j} = \ket{d_{-1}}$:
	\begin{equation}
		\begin{array}{c}
			\epsilon_{0,-2} \ket{E_{0,-2}} + \epsilon_{0,-1} \ket{E_{0,-1}} = 0 \\
			\Downarrow \\
			\epsilon_{0,-2} = \epsilon_{0,-1} = 0.
		\end{array}
	\end{equation}

	\item For $\ket{j} = \ket{d_2}$:
	\begin{equation}
		\begin{array}{c}
			\epsilon_{0,1} \ket{E_{0,1}} + \epsilon_{0,2} \ket{E_{0,2}} = 0 \\
			\Downarrow \\
			\epsilon_{0,1} = \epsilon_{0,2} = \epsilon_{0,3} = 0.
		\end{array}
	\end{equation}

	\item For $\ket{j} = \ket{d_3}$:
	\begin{equation}
		\begin{array}{c}
			\underset{=0}{\epsilon_{0,2}} \ket{E_{0,2}} + \underset{=0}{\epsilon_{0,3}} \ket{E_{0,3}} = 0.
		\end{array}
	\end{equation}
\end{itemize}

Now, consider Alice's state $\ket{\psi}_{\sA} = \ket{\fone} = \ket{t'_1}$. The erroneous states the must be avoided are as follows:
\begin{equation}
	J_{\err} = J_0 \cup J_{\inv} = \left\{ \ket{s_{-1}}, \ket{s_0}, \ket{s_3}, \ket{d_{-1}}, \ket{d_0}, \ket{d_3} \right\}.
\end{equation}

\begin{itemize}
	\item For $\ket{j} = \ket{s_{-1}}$:
	\begin{equation}
		\epsilon_{1,-2} \ket{E_{1,-2}} - \epsilon_{1,-1} \ket{E_{1,-1}} = 0.
	\end{equation}

	\item For $\ket{j} = \ket{s_0}$:
	\begin{equation}
		\begin{array}{c}
			\epsilon_{1,-1} \ket{E_{1,-1}} - \epsilon_{1,0} \ket{E_{1,0}} = 0 \\
			\Downarrow \\			
			\epsilon_{1,-2} = \epsilon_{1,-1} = \epsilon_{1,0}.
		\end{array}
	\end{equation}

	\item For $\ket{j} = \ket{s_3}$:
	\begin{equation}
		\epsilon_{1,2} \ket{E_{1,2}} - \epsilon_{1,3} \ket{E_{1,3}} = 0.
	\end{equation}

	\item For $\ket{j} = \ket{d_{-1}}$:
	\begin{equation}
		\begin{array}{c}
			\epsilon_{1,-2} \ket{E_{1,-2}} + \epsilon_{1,-1} \ket{E_{1,-1}} = 0 \\
			\Downarrow \\
			\epsilon_{1,-2} = \epsilon_{1,-1} = \epsilon_{1,0} = 0.
		\end{array}
	\end{equation}

	\item For $\ket{j} = \ket{d_0}$:
	\begin{equation}
		\underset{=0}{\epsilon_{1,-1}} \ket{E_{1,-1}} + \underset{=0}{\epsilon_{1,0}} \ket{E_{1,0}} = 0 .
	\end{equation}

	\item For $\ket{j} = \ket{d_3}$:
	\begin{equation}
	\label{eqn:defense_z_1_2}
		\begin{array}{c}
			\epsilon_{1,2} \ket{E_{1,2}} + \epsilon_{1,3} \ket{E_{1,3}} = 0 \\
			\Downarrow \\
			\epsilon_{1,2} = \epsilon_{1,3} = 0.
		\end{array}
	\end{equation}
\end{itemize}

From Eqs.~\eqref{eqn:defense_z_0_-2}--\eqref{eqn:defense_z_1_2},
we notice that the only non-zero coefficients left are $\epsilon_{0,0}$ and $\epsilon_{1,1}$.
Thus, our attack must satisfy:
\begin{equation}
\label{eqn:defense_almost_there}
	\begin{array}{rcl}
		U_{\sE} \ket{0}_{\sE} \ket{t'_0}_{\sA} &=& \epsilon_{0,0}\ket{E_{0,0}}_{\sE} \ket{t'_0}, \\
		U_{\sE} \ket{0}_{\sE} \ket{t'_1}_{\sA} &=& \epsilon_{1,1}\ket{E_{1,1}}_{\sE} \ket{t'_1}. \\
	\end{array}
\end{equation}

\paragraph{Hadamard Basis}

Let us consider $\ket{\psi}_{\sA} = \ket{+} = \frac{1}{\sqrt{2}}(\ket{t'_0} + \ket{t'_1})$.
We must avoid the states:
\begin{equation}
	J_{\err} = J_{1} \cup J_{\inv} = \left\{ \ket{s_{-1}}, \ket{d_{-1}}, \ket{s_1}, \ket{s_3}, \ket{d_3} \right\}.
\end{equation}

For $\ket{j} = \ket{s_1}$, we get:
\begin{equation}
\label{eqn:defense_x}
	\begin{array}{c}
		\sum_k \epsilon_{0,k}\beta_{k,d_1}\ket{E_{0,k}} + \sum_k \epsilon_{1,k}\beta_{k,d_1}\ket{E_{1,k}} = 0 \\
		\Downarrow \\
		\frac{i}{\sqrt{2}}(-\epsilon_{0,0}\ket{E_{0,0}} + \underset{=0}{\epsilon_{0,1}}\ket{E_{0,1}}) + \frac{i}{\sqrt{2}}(-\underset{=0}{\epsilon_{1,0}}\ket{E_{1,0}} + \epsilon_{1,1}\ket{E_{1,1}}) = 0 \\
		\Downarrow \\
		\epsilon_{0,0}\ket{E_{0,0}} = \epsilon_{1,1}\ket{E_{1,1}}.
	\end{array}
\end{equation}

Applying the result of Eq.~\eqref{eqn:defense_x} to Eq.~\eqref{eqn:defense_almost_there}, we get that 
the only attack Eve can apply without introducing noise is the trivial attack:
\begin{equation}
	\begin{array}{c}
		U_{\sE} \ket{0}_{\sE} \ket{\fzero}_{\sA} = \ket{\phi}_{\sE} \ket{\fzero}_{\sA}, \\
		U_{\sE} \ket{0}_{\sE} \ket{\fone}_{\sA} = \ket{\phi}_{\sE} \ket{\fone}_{\sA} \\
		\Downarrow \\
		U_{\sE} \ket{0}_{\sE} \ket{\psi}_{\sA} = \ket{\phi}_{\sE} \ket{\psi}_{\sA}.
	\end{array}
\end{equation}
Which is completely independent of Alice's input state. Applying the additional constraints derived from the Hadamard basis is unnecessary at this point.

We have shown there is no attack that satisfies the requirements defined at the start of the subsection (single-photon states,
each input state attacked separately, no added vulnerabilities, zero added noise) and gains information on the shared key.
\qed

\section{Computation of Bright Illumination Attack Using Reversed-Space}

\label{app:bi_rsa_exploit}
In this appendix, we show the full computation of the Reversed-Space Attack discussed in Section~\ref{subsec:bi_rsa}, beginning from the behavior of the implementation that was discovered through Quantum Fuzzing in Section~\ref{subsec:bi_fuzz} and resulting in the original attack described in Section~\ref{subsec:bi_explanation}.
Our modeling of the device will rely \textit{only} on the information learned on the device from fuzzing, and on the fact that the device employs a single unitary transformation independently of the basis it wishes to measure.

\subsection*{Bob's Measured Space}
The vulnerabilities shown in Section~\ref{subsec:bi_fuzz} show that Bob's device has an unintended configuration that can be triggered by Eve (the linear mode of the APDs), and that in this configuration, his device suffers from a collection of Measurement Space Vulnerabilities and Interpretation Vulnerabilities (as defined in Section~\ref{subsubsec:qkd_vulnerabilities}). We will now formally describe these vulnerabilities.

For a polarization-based BB84 state $\ket{\psi}$, let us use $\ket{\psi}^\mathrm{bright}$ to denote the high-photon-number state of the same polarization.

Let us denote Bob's unitary transformation on the input as $U_{\sB}$. Note that there is only one transformation $U_{\sB}$ (that is, $m = 1$, in the notation of Section~\ref{subsec:RSA_eve_and_bob}), because Bob uses a passive basis choice.

We denote the input states \textit{after} the application of $U_{\sB}$ in the following way:

\begin{equation}
    \begin{array}{rclrcl}
        \ket{\psi_0}_{\sB} &\triangleq& U_{\sB}\ket{0}^\mathrm{bright}, & \quad
        \ket{\psi_1}_{\sB} &\triangleq& U_{\sB}\ket{1}^\mathrm{bright}, \\[1.2ex]
        \ket{\psi_2}_{\sB} &\triangleq& U_{\sB}\ket{+}^\mathrm{bright}, & \quad
        \ket{\psi_3}_{\sB} &\triangleq& U_{\sB}\ket{-}^\mathrm{bright}. \\
    \end{array}
\end{equation}

Note that, according to the definition of high-photon-number states with diagonal polarization (see Appendix~\ref{app:quantum_optics}), the overlap between ``bright Hadamard'' states and ``bright computational'' states is given, without loss of generality, by:

\begin{equation}
	\bra{+}^\mathrm{bright} \ket{0}^\mathrm{bright} = \left(\frac{1}{\sqrt{2^k}} \sum_{l=0}^k \sqrt{\binom{k}{l}} \bra{l, k-l}_{\mathrm{H},\mathrm{V}} \right) \ket{0,k}_{\mathrm{H},\mathrm{V}} = \frac{1}{\sqrt{2^k}} = O \left(2^{-\frac{k}{2}}\right).
\end{equation}

Thus, if the photon number $k$ is high enough, we can approximate
$$\ket{\fzero}^\mathrm{bright}, \ket{\fone}^\mathrm{bright} \in \spn\left\{\ket{+}^\mathrm{bright}, \ket{-}^\mathrm{bright}\right\}^\bot.$$
and vice versa.

Therefore, the set $\left\{\ket{\fzero}^\mathrm{bright}, \ket{\fone}^\mathrm{bright}, \ket{+}^\mathrm{bright}, \ket{-}^\mathrm{bright}\right\}$ is orthonormal, and since $U_{\sB}$ is unitary, so is the set $\left\{ \ket{\psi_0}_{\sB}, \ket{\psi_1}_{\sB}, \ket{\psi_2}_{\sB}, \ket{\psi_3}_{\sB} \right\}$.
We thus model the Hilbert space that Bob measures as:
\begin{equation}
    \mathcal{H}^{\sB} = \spn\left\{ \ket{\psi_0}_{\sB}, \ket{\psi_1}_{\sB}, \ket{\psi_2}_{\sB}, \ket{\psi_3}_{\sB} \right\}.
\end{equation}

Note that, since this is a BB84 implementation with passive basis choice, Bob's choice of basis is performed by the device itself.
From the Quantum Fuzzing procedure, we know Bob's measurement interpretations (as defined in Section~\ref{subsec:RSA_oblivious}) in both bases:

\begin{itemize}
\item In the computational basis:

\begin{equation}
    J_0 = \{\ket{\psi_0}_{\sB}\}, \quad
    J_1 = \{\ket{\psi_1}_{\sB}\}.
\end{equation}

Since the states $\ket{\psi_2}_\sB$ and $\ket{\psi_3}_\sB$ always cause a Hadamard basis measurement, they can never be interpreted as part of a computational basis measurement, and can be omitted from consideration in this basis.

\item In the Hadamard basis:

\begin{equation}
    J_0 = \{\ket{\psi_2}_{\sB}\}, \quad
    J_1 = \{\ket{\psi_3}_{\sB}\}.
\end{equation}

Similarly, since the states $\ket{\psi_0}_\sB$ and $\ket{\psi_1}_\sB$ always cause a computational basis measurement, they can never be interpreted as part of a Hadamard basis measurement, and can be omitted from consideration in this basis.

\end{itemize}

\subsection*{Eve's Attack}
When reversing the space that Bob measures through Bob's unitary, we unsurprisingly get the space spanned by the four states we found by Quantum Fuzzing:

\begin{equation}
    \mathcal{H}^{\sP} = \spn \left\{ U_{\sB}^\dagger \ket{\psi_i}_{\sB} \;\big|\; i=0,1,2,3\right\}
    = \spn \left\{\ket{\fzero}^\mathrm{bright}, \ket{\fone}^\mathrm{bright}, \ket{+}^\mathrm{bright}, \ket{-}^\mathrm{bright}\right\}.
\end{equation}

Thus, Eve's general form of attack is:

\begin{equation}
	\begin{array}{rclclc}
		U_{\sE} \ket{\fzero}_{\sE} \ket{\fzero}_{\sA} &=& \epsilon_{0,0}\ket{E_{0,0}}\ket{\fzero}^\mathrm{bright} &+& \epsilon_{0,1}\ket{E_{0,1}}\ket{\fone}^\mathrm{bright} &+ \\
		& & \epsilon_{0,2}\ket{E_{0,2}}\ket{+}^\mathrm{bright} &+& \epsilon_{0,3}\ket{E_{0,3}}\ket{-}^\mathrm{bright}, \\
        \\
		U_{\sE} \ket{\fzero}_{\sE} \ket{\fone}_{\sA} &=& \epsilon_{1,0}\ket{E_{1,0}}\ket{\fzero}^\mathrm{bright} &+& \epsilon_{1,1}\ket{E_{1,1}}\ket{\fone}^\mathrm{bright} &+ \\
		& & \epsilon_{1,2}\ket{E_{1,2}}\ket{+}^\mathrm{bright} &+& \epsilon_{1,3}\ket{E_{1,3}}\ket{-}^\mathrm{bright}.

	\end{array}
\end{equation}

\subsection*{Attack Constraints}
The constraints, as defined in Section~\ref{subsec:RSA_oblivious}, are:
\begin{equation}
    \begin{array}{c}
         \forall \ket{\psi}_{\sA} = \sum_i \alpha_i \ket{i}_{\sA},\quad \forall \ket{j} \in J_{\err}:  \\[1.5ex]
         \sum_{i,k}\alpha_{i} \epsilon_{i,k}\beta_{k,j} \ket{E_{i,k}}=0,
    \end{array}
\end{equation}
where $ U_{\sB} = \sum_{k,j} \beta_{k,j} \ket{j}\bra{k} $  is Bob's unitary transformation, together with orthonormality conditions on Eve's output states.

In this case, Bob's transformation is defined as $\beta_{k,j} = \delta_{k,j}$ (Kronecker delta). Therefore, the condition for our attack is: $\sum_{i,k}\alpha_{i} \epsilon_{i,k}\delta_{k,j} \ket{E_{i,k}}=0$, which is easily simplified to:
\begin{equation}
    \sum_i\alpha_{i} \epsilon_{i,j} \ket{E_{i,j}}=0,
\end{equation}   
for all $\ket{j} \in J_{\err}$.
Let us apply the constraints to each measurement basis and each input state separately.

\subsubsection*{Computational Basis}

\begin{itemize}
	\item For $\ket{\fzero}_{\sA}$: $\vec{\alpha} = (1,0)$. 
	$$\forall \ket{j} \in J_{\err}:\; \epsilon_{0,j}\ket{E_{0,j}} = 0. $$
	Since  $J_{\err} = J_1 = \{\ket{\psi_1}_{\sB}\}$:
	\begin{equation}
		\label{eqn:bi_01}
		\epsilon_{0,1}\ket{E_{0,1}} =0.
	\end{equation}

	\item For $\ket{\fone}_{\sA}$: $\vec{\alpha} = (0,1)$.
	$$\forall \ket{j} \in J_{\err}:\; \epsilon_{1,j}\ket{E_{1,j}} = 0. $$
	Since $J_{\err} = J_0 = \{\ket{\psi_0}_{\sB}\}$:
	\begin{equation}
		\label{eqn:bi_10}
		\epsilon_{1,0}\ket{E_{1,0}} = 0.
	\end{equation}
\end{itemize}

\subsubsection*{Hadamard Basis}

\begin{itemize}
	\item For $\ket{+}_{\sA}$: $\vec{\alpha} = \left(\frac{1}{\sqrt{2}}, \frac{1}{\sqrt{2}}\right)$.
	$$\forall \ket{j} \in J_{\err}:\; \epsilon_{0,j}  \ket{E_{0,j}} +  \epsilon_{1,j} \ket{E_{1,j}} = 0. $$
        Since  $J_{\err} = J_1 = \{\ket{\psi_3}_{\sB}\}$, we get:
	\begin{equation}
		\varepsilon_{0,3}\ket{E_{0,3}} + \varepsilon_{1,3}\ket{E_{1,3}} = 0.
	\end{equation}

	\item For $\ket{-}_{\sA}$: $\vec{\alpha} = \left(\frac{1}{\sqrt{2}}, \frac{-1}{\sqrt{2}}\right)$. 
    $$\forall \ket{j} \in J_{\err}:\quad \epsilon_{0,j}  \ket{E_{0,j}} - \epsilon_{1,j} \ket{E_{1,j}} = 0. $$
    
    Since  $J_{\err} = J_0 = \{\ket{\psi_2}_{\sB}\}$, we get:
    \begin{equation}
        \label{eqn:bi_02_12}
        \epsilon_{0,2}\ket{E_{0,2}} - \epsilon_{1,2}\ket{E_{1,2}} = 0.
    \end{equation}
\end{itemize}

Combining Eqs.~\eqref{eqn:bi_01}--\eqref{eqn:bi_02_12}, our generic attack can be written as:

\begin{equation}
	\begin{array}{rcccccl}
		U_{\sE} \ket{0}_{\sE}\ket{0}_{\sA} &=&
        \epsilon_{0,0}\ket{E_{0,0}}\ket{0}^\mathrm{bright} &+& 
        \epsilon_{0,2}\ket{E_{0,2}}\ket{+}^\mathrm{bright} &+&
        \epsilon_{0,3}\ket{E_{0,3}}\ket{-}^\mathrm{bright}, \\
		U_{\sE} \ket{0}_{\sE}\ket{1}_{\sA} &=&
        \epsilon_{1,1}\ket{E_{1,1}}\ket{1}^\mathrm{bright} &+&
        \epsilon_{1,2}\ket{E_{1,2}}\ket{+}^\mathrm{bright} &+&
        \epsilon_{1,3}\ket{E_{1,3}}\ket{-}^\mathrm{bright},
	\end{array}
\end{equation}
with the constraints:
\begin{equation}
	\begin{array}{rcl}
 	\epsilon_{0,2}\ket{E_{0,2}} - \epsilon_{1,2}\ket{E_{1,2}} &=& 0,	\\
        \epsilon_{0,3}\ket{E_{0,3}} + \epsilon_{1,3}\ket{E_{1,3}} &=& 0, \\
        ||U_{\sE} \ket{0}_{\sE} \ket{0}_{\sA}||^2 &=& 1, \\
        ||U_{\sE} \ket{0}_{\sE} \ket{1}_{\sA}||^2 &=& 1, \\
        (U_{\sE} \ket{0}_{\sE} \ket{0}_{\sA})^\dagger  (U_{\sE} \ket{0}_{\sE} \ket{1}_{\sA}) &=& 0.
	\end{array}
\end{equation}

This can be algebraically reduced to the following form:
\begin{equation}
	\begin{array}{rcccccl}
		U_{\sE} \ket{0}_{\sE}\ket{0}_{\sA} &=&
        p\ket{E_0}\ket{0}^\mathrm{bright} &+&
        q\ket{E_2}\ket{+}^\mathrm{bright} &+&
        q\ket{E_3}\ket{-}^\mathrm{bright}, \\
            U_{\sE} \ket{0}_{\sE}\ket{1}_{\sA} &=&
        p\ket{E_1}\ket{1}^\mathrm{bright} &+&
        q\ket{E_2}\ket{+}^\mathrm{bright} &-&
        q\ket{E_3}\ket{-}^\mathrm{bright}, \\
        
	\end{array}
\end{equation}
where $p$ and $q$ are non-negative real numbers (because without loss of generality, phase can be incorporated into Eve's ancillary states), with the constraint: 
\begin{equation}
    p^2 + 2q^2 = 1.
\end{equation}

This concludes our construction of the attack. \qed

\section{Examples of Attacks on QKD}
\label{app:attack_list}
This appendix lists QKD attacks and attack families used in Section~\ref{sec:applicability},
as well as high-level details on each attack.

\subsection{List of Individual Attacks}
\begin{itemize}

    \item \textbf{Large Pulse Attack}~\cite{vakhitov_large_2001}
    
    This attack is discussed in Section~\ref{subsec:side_channel_examples}.

    \item \textbf{Photon-Number Splitting (PNS) Attack}~\cite{PNS,PNS_eurocrypt}
    
    This attack is discussed in Section~\ref{subsubsec:qkd_vulnerabilities} of our work. Additional details appear in Section~2.5.1 of~\cite{bsi}.

    \item \textbf{Injection-Locking Attack}~\cite{Pang2020}

    This attack is discussed in Section~\ref{subsec:side_channel_examples} of our work. Additional details appear in Table~4.35 of~\cite{bsi}.
    
    \item \textbf{Time-Shift Attack}~\cite{qi_time-shift_2006}

    This attack is applicable in QKD receivers that use different photo-detectors to detect different states, and the detection windows of the detectors are not identical.
    Eve selectively makes Alice's signal arrive earlier or later than the original detection time, making only one detection result possible. This attack does not rely on Eve measuring Alice's signal to generate an appropriate state to send to Bob, which makes the attack highly practical.

    Further details can be found in Table~4.5 of~\cite{bsi}.

    \item \textbf{Trojan Pony Attack}~\cite{gottesman_security_2004,liss_practice_2020}
    
    This attack is initially defined in~\cite{gottesman_security_2004} as one built under a two-adversary model.
    In this model, an additional adversary named Fred is inside Bob's lab, and can get classical information
    from Eve to affect Bob's device in a limited way. Fred cannot communicate back to Eve.
    In this setup, Bob's detector is not fully efficient, and Fred has control over when to fail Bob's detector.
    When Fred knows that Eve and Bob chose mismatching bases, he drops Eve's photon.
    
    A special case of this attack, given in~\cite{gottesman_security_2004}, is a detector
    which behaves identically for pulses with a single photon and those with a high number of photons,
    and Bob treats double-click events as losses.
    In their work, \cite{liss_practice_2020} show that their implementation of this attack can be built using Reversed-Space.
\end{itemize}

\subsection{List of Attack Families}

\begin{itemize}
    \item \textbf{Bright Illumination Attacks}~\cite{lydersen_blinding_commercial_2010,sague_blinding_pulsed_2011,lydersen_blinding_thermal_2010}
    
    The pulsed variant of this attack~\cite{sague_blinding_pulsed_2011} is analyzed in Section~\ref{sec:bi_analysis}.
    Different variants appear in tables 4.12-4.14 and 4.17-4.20 of~\cite{bsi}.

    \item \textbf{Faked States Attacks}~\cite{makarov_faked_2005,makarov_effects_2006,makarov_faked_2008}
    
    This is a general family of intercept-resend attacks where Eve sends malicious signals to Bob,
    which cause him to get a result of her choice, as discussed in Section~\ref{subsubsec:qkd_exploits} of our work. Additional details appear in Section 2.5.2.1 of~\cite{bsi}.
    
    \item \textbf{Fixed Apparatus Attacks}~\cite{boyer_fixedapp_2014}
    
    These attacks are applicable against all QKD implementations where Bob's
    choice of basis is not active --- for example, when a beam splitter
    is used for replacing Bob's true choice of basis.
    
    In some cases of the Fixed-Apparatus attack, Bob's passive choice alone
    performs an enlargement of $\mathcal{H}^{\sB}$ which Eve can use to build a Reversed-Space Attack.
    
    In its general case, this attack model also considers a situation where
    Eve gains access to Bob's ancillary space and can manipulate Bob's
    ancillary state.
    Using this power, Bob's effective measured space is enlarged for Eve
    (to be $\mathcal{H}^{\sB} \otimes \mathcal{H}^{\sanc}$ instead of $\mathcal{H}^{\sB}$) and she can build a Reversed-Space Attack.

    \item \textbf{Trojan Horse Attacks}~\cite{Gisin2002,gisin_trojan-horse_2006}
    
    Trojan Horse attacks send light pulses into Alice or Bob's devices, and study
    secret aspects of their configuration from the back-scattered light.
    Details appear in Table~4.48 of~\cite{bsi}.
    These attacks are a general case of the Large Pulse attack, as defined by~\cite{vakhitov_large_2001}. The Large Pulse attack specifically probes devices by sending a high-intensity light pulse at a specific window, whereas Trojan Horse attacks are relatively generic.
    Note that Lo and Chau
    defined a different notion of ``Trojan Horse'' attacks~\cite{lo_unconditional_1999,lo_proof_2001}, but it is rarely used.

    \item \textbf{Reversed-Space Attacks}~\cite{gelles_security_2012,gelles_reversed_2016}
    
    These attacks are discussed in Section~\ref{sec:RSA}.

    \item \textbf{Detector Efficiency Mismatch Attacks}~\cite{makarov_faked_2005,makarov_effects_2006}
    
    This attack utilizes a receiver that uses several detectors in order to measure the incoming states, and their sensitivities are not completely identical. Eve uses the non-overlapping parts of the windows and sends fake states that force a specific state/basis measurement. As such, it is a special case of the Faked-States attack family.
    Details appear in Table~4.5 of~\cite{bsi}.
    
\end{itemize}

\end{document}